# Electric-field control of spin-orbit torque in a magnetically doped topological insulator

Yabin Fan[1†*], Xufeng Kou[1†], Pramey Upadhyaya[1†], Qiming Shao[1], Lei Pan[1], Murong Lang[1], Xiaoyu Che[1], Jianshi Tang[1], Mohammad Montazeri[1], Koichi Murata[1], Li-Te Chang[1], Mustafa Akyol[1], Guoqiang Yu[1], Tianxiao Nie[1], Kin L. Wong[1], Jun Liu[3], Yong Wang[3], Yaroslav Tserkovnyak[2] and Kang L. Wang[1*]

[1]Department of Electrical Engineering, University of California, Los Angeles, California 90095, USA

[2]Department of Physics and Astronomy, University of California, Los Angeles, California 90095, USA

[3]Center of Electron Microscopy and State Key Laboratory of Silicon Materials, School of Materials Science and Engineering, Zhejiang University, Hangzhou 310027, China

[†]These authors contributed equally to this work.

*To whom correspondence should be addressed. E-mail: yabin@seas.ucla.edu; wang@seas.ucla.edu

**Electric-field manipulation of magnetic order has proved of both fundamental and technological importance in spintronic devices. So far, electric-field control of ferromagnetism, magnetization and magnetic anisotropy has been explored in various magnetic materials, but the efficient electric-field control of spin-orbit torque (SOT) still remains elusive. Here, we report the effective electric-field control of a giant SOT in a Cr-doped topological insulator (TI) thin film using a top-gate FET structure. The SOT strength can be modulated by a factor of 4 within the accessible gate voltage range, and it shows strong correlation with the spin-polarized surface current in the film. Furthermore, we demonstrate the magnetization switching by scanning gate voltage with constant current and in-plane magnetic field applied in the film. The effective electric-field control of SOT and the giant spin-torque efficiency in Cr-doped TI may lead to the development of energy-efficient gate-controlled spin-torque devices compatible with modern field-effect semiconductor technologies.**

Electric-field control of magnetization dynamics is essential for realizing high-performance, low-power spintronic memory and logic devices. During recent years, enormous progress has been made in this regard. Among notable achievements are the electric-field control of ferromagnetism in magnetic semiconductors (*e.g.*, Mn-doped InAs, GaAs) through modulation of the carrier (hole) concentration [1,2], electric-field manipulation of magnetization direction and/or reversal by use of multiferroics [3-5], and significant voltage-controlled magnetic anisotropy change in ultrathin ferromagnet/oxide junctions (*e.g.*, Fe/MgO) [6,7]. On the other hand, for another method of manipulating magnetic moment --- current-induced spin-orbit torque (SOT) [8,9], which is important in practical devices, the electric gate control is just beginning to be probed [10,11]. Thus, it becomes highly desirable to study the efficient electric-field control of SOT in novel magnetic structures, which may allow for the manipulation of magnetization in a fashion compatible with modern field-effect semiconductor devices. Recently, topological insulator (TI) [12-14] -based magnetic structures have drawn much attention due to the giant spin-torque efficiency potentially applicable for spintronic devices [15-17]. Compared with conventional heavy metal/ferromagnet heterostructures (HMFHs) where the spin-Hall effect in the heavy metal and/or the Rashba-Edelstein effect at the interface are crucial for generating the spin torque [18-22], TIs exhibit surface spin-momentum locked Dirac fermions [23-30] which are expected to be more efficient [31,32]. Indeed, a giant current-induced SOT in TI-based magnetic structures has been reported in several recent experiments [15-17] and related spin pumping/inverse spin Hall effect was also investigated in these structures [33-36].

Another important feature of TI is its bulk behavior as a semiconductor which allows effective electric-field manipulation of the surface carrier density and type [37-39], as opposed to heavy metals. Thus, the long-sought-after electric-field control of SOT may be potentially realized in TI-based magnetic structures. In this paper, we report the gate electric-field control of SOT in the nominally uniformly Cr-doped TI (Cr-TI, for short) thin film in the Au(electrode)/Al$_2$O$_3$/Cr-TI/GaAs(substrate) structure. Using the low-frequency harmonic method [9,15,40], we show that the SOT strength can be modulated by a factor of 4 by gate tuning within the accessible voltage range, which is almost two orders of magnitude larger than that reported in HMFHs [10]. The effective gate control, as discussed later in our analysis, also shows strong correlation with the net spin-polarized surface current originating from the topological surface states (SS) in the film. Furthermore, we demonstrate that the magnetization can be switched by scanning gate voltage with constant current and in-plane magnetic field applied in the Cr-TI film, pointing towards practical device applications such as electric-field controlled magnetic memory switch compatible with modern field-effect semiconductor devices. Thus, the electric-field control is promising as another approach besides the lateral current to harness the giant SOT in the TI-based magnetic structures, which can potentially add new functionalities to spin-torque devices such as simultaneous memory and logic functions.

**Results**

**Magnetization switching through current-induced SOT in the $Al_2O_3$/Cr-TI/GaAs(substrate) structure.** Before introducing the top gate electrode, we first briefly examine an $Al_2O_3$/Cr-TI/GaAs(substrate) structure to investigate the basic material properties and the intrinsic current-induced SOT. As shown in Fig. 1a, a 7-quintuple-layer (QL) uniformly Cr-doped $Cr_{0.16}(Bi_{0.50}Sb_{0.42})_2Te_3$ thin film was epitaxially grown on an insulating GaAs (111)B substrate using molecular beam epitaxy (MBE) [37], and a 20-nm-thick high-$k$ $Al_2O_3$ dielectric layer was deposited on top to protect the Cr-TI film [41]. Inside the Cr-TI layer, the blue arrows indicate the Cr dopant elements which are uniformly distributed [37,38] (see Supplementary Section 1). It is known that in the Cr-TI film, ferromagnetism can be developed through bulk van-Vleck mechanism [38] and/or surface carrier mediated magnetism [37,38,42] at low temperature. To probe the magnetic properties, the thin film was patterned into micrometer scale Hall bar structure (see Methods) [39,41], as shown in Fig. 1b inset, which enables measurement of both transverse (Hall) and longitudinal resistances. In this Hall bar configuration, we define the current flowing from the left to the right (*i.e.*, along *y*–direction) as the positive current; $V_H$ measures the Hall voltage and $V_L$ measures the longitudinal voltage. Both the out-of-plane Hall resistance $R_H$ and the longitudinal magneto-resistance $R_L$ measured at 1.9K are plotted in Fig. 1b, as functions of the out-of-plane magnetic field. The nearly square-shape $R_\text{H}$ hysteresis loop together with the butterfly-shape $R_\text{L}$ demonstrate the pronounced magnetism [37,38] in the Cr-TI film with a magnetic easy axis along the out-of-plane direction (*i.e.*, along *z*-direction). Given the low Curie temperature of the film ($T_\text{C} = 11$K as discussed in Supplementary Section 1), in the following, all the transport experiments are performed at 1.9K unless otherwise stated.

In this Al$_2$O$_3$/Cr-TI/GaAs structure, even though the uniform doping inside the Cr-TI layer respects the inversion symmetry along the *z*-direction, the structural asymmetry due to Al$_2$O$_3$ and GaAs can induce different interfacial properties with the Cr-TI layer, as illustrated in Fig. 1a, where the bottom topological SS carrier density is estimated to be higher than that of the top SS (more evidences are provided in the Discussions section and Supplementary Section 3). By passing a charge current through the Cr-TI layer, the spin-momentum locked SS carriers on the two interfaces can produce non-equal amount of torques to the Cr dopant magnetization, and thus a non-zero net SOT is expected. In Fig. 1a, on top of the Al$_2$O$_3$/Cr-TI/GaAs structure we depict the total magnetization $\boldsymbol{M}$, the external magnetic field $\boldsymbol{B}_\text{ext}$ and various torques: $\boldsymbol{\tau}_\text{ext}$ exerted by $\boldsymbol{B}_\text{ext}$, $\boldsymbol{\tau}_\text{K}$ exerted by the perpendicular anisotropy field $\boldsymbol{B}_\text{K}$, and the net current-induced SOT, $\boldsymbol{\tau}_\text{SO}$, for a DC current $I_\text{dc}$ flowing along the $-y$ direction in the film. Here, $\boldsymbol{M}$ and $\boldsymbol{B}_\text{ext}$ are both in the *yz* plane, with $\theta_\text{M}$ and $\theta_\text{B}$ being the respective polar angles relative to the *z*-axis; $\boldsymbol{B}_\text{K} = K\cos\theta_\text{M}\hat{\boldsymbol{z}}$, and $K$ is the out-of-plane anisotropy coefficient (see Supplementary Section 4); $\boldsymbol{\tau}_\text{SO} = -\gamma \boldsymbol{M} \times \boldsymbol{B}_\text{SO}$, where $\gamma$ is the gyromagnetic ratio and $\boldsymbol{B}_\text{SO} = I_\text{dc}\lambda_\text{SO}\hat{\boldsymbol{x}} \times \boldsymbol{m}$ is the current-induced effective spin-orbit field [15]; $\lambda_\text{SO}$ is the coefficient characterizing the spin-orbit coupling strength in the system and $\boldsymbol{m}$ is the unit vector denoting the magnetization direction. In this structure, $\boldsymbol{\tau}_\text{SO}$ is the so-called damping-like SOT which is the dominant term [15] and consistent with the bottom surface spin-polarized current originating from the topological SS [23-30]. Thus, the equilibrium state of $\boldsymbol{M}$ can be achieved by balancing the torques: $\boldsymbol{\tau}_\text{ext} + \boldsymbol{\tau}_\text{K} + \boldsymbol{\tau}_\text{SO} = 0$. In order to demonstrate the current-induced $\boldsymbol{\tau}_\text{SO}$ in the structure, we carried out the ($B_y$-fixed, $I_\text{dc}$-dependent) magnetization switching experiment. Indeed, the out-of-plane component of the magnetization, $M_Z$ (manifested by the anomalous Hall effect (AHE) resistance $R_{AHE}$), can be successfully switched by scanning the in-plane DC current in the presence of the fixed in-plane magnetic fields $\boldsymbol{B_y} = B_y \hat{\boldsymbol{y}}$ with $B_y = \pm 2$T, as shown in Fig. 1c. The switching is hysteretic and agrees with the definitions of $\boldsymbol{\tau}_{SO}$ and $\boldsymbol{B}_{SO}$ in Fig. 1a. The correlated ($I_{dc}$-fixed, $B_y$-dependent) switching experiment can be found in Supplementary Section 1. We summarize the switching behaviors as a phase diagram in Fig. 1d, from which we see that $M_Z$ can be switched using only tens of micro-amperes of DC current, indicating that even in this uniformly doped Al$_2$O$_3$/Cr-TI/GaAs structure, the current-induced SOT can be significant.

We furthermore carried out the low-frequency second harmonic experiment [9,15,40] to quantitatively measure the strength of the current-induced SOT. By sending an AC current into the Cr-TI layer, the current-induced alternating $\boldsymbol{B_{SO}}$ field causes the magnetization $\boldsymbol{M}$ to oscillate around its equilibrium position, which gives rise to the second harmonic AHE resistance, $R_{AHE}^{2\omega}$. In Fig. 1e, we plot the measured $R_{AHE}^{2\omega}$ as a function of the in-plane magnetic field for input AC current with different *rms* values from 2 μA to 10 μA. The low frequency $\omega$ utilized is 15.8 Hz. In this case, the effective $\boldsymbol{B_{SO}}$ field is pointing along $\hat{\boldsymbol{z}}$ or $-\hat{\boldsymbol{z}}$ direction, and for convenience we label it as $\boldsymbol{B_{SO}} = B_{SO} \sin(\omega t) \hat{\boldsymbol{z}}$. In the large field region ($|B_y| \gg K$), as shown in Fig. 1e, $R_{AHE}^{2\omega}$ can be fitted using the formula (see Supplementary Section 5) [15],

$$R_{AHE}^{2\omega} = -\frac{1}{2} R_A B_{SO} / (|B_y| - K) \tag{1}$$

where $R_A$ is the out-of-plane saturation AHE resistance (see Supplementary Section 4). The $1/(|B_y| - K)$ scaling relation in the large field region is consistent with the argument on SOT-induced magnetization oscillation around its equilibrium position. In Fig. 1f, we plot the obtained effective spin-orbit field $B_{SO}$ as a function of $I_{ac}$ (*rms* value) for both the $B_y > 0$ and the $B_y < 0$ cases. The directions of the obtained $\boldsymbol{B_{SO}}$ are consistent with the switching experiments and the effective field versus current ratio is $\frac{|B_{SO}|}{I_{ac}^{peak}} =$ 6.7 mT/μA by linear fitting. If we tentatively assume the current distribution inside the Cr-TI layer is uniform, then the corresponding spin-torque ratio [9,15,16,18] can be estimated as, $\vartheta_{ST} = \frac{2eM_S B_{SO} w t_{Cr-TI}^2}{\hbar I_{ac}^{peak}} = 81$ ($e$ is the electron charge, $\hbar$ is the reduced Planck constant, $M_S = 8.5 \text{ emu/cm}^3$ is the magnetization magnitude, $w$ and $t_{Cr-TI}$ are the width and thickness of the Cr-TI layer), which again demonstrates that even in the uniformly doped $Al_2O_3$/Cr-TI/GaAs structure, due to the two non-balanced interfaces, the current-induced SOT can be large with the corresponding spin-torque efficiency much higher than that observed in HMFHs [9,40] (Note: a more accurate $\vartheta_{ST}$ will be provided by considering the more realistic current distribution inside the Cr-TI layer in the Discussions section).

**Electric-field effect on material properties in the Au(electrode)/$Al_2O_3$/Cr-TI/GaAs(substrate) structure.** The current-induced SOT arising from the non-balanced topological SS carrier distributions at the two interfaces in the $Al_2O_3$/Cr-TI/GaAs structure inspires us to pursue the external gate electric-field control of SOT since the external gate voltage can effectively tune the surface carrier density and type in the Cr-TI layer [37,38], and consequently manipulate the current-induced SOT strength. Following this idea, we deposited an Au electrode as a top gate on the $Al_2O_3$/Cr-TI/GaAs structure to form a gate controllable Hall bar device, as shown in Fig. 2a. A gate voltage of $V_g$ can be applied between the top gate and the source contact. Before measuring the SOT under different $V_g$, we first examine the gate electric-field effect on the characteristics of the Cr-TI film. Figure 2b shows the longitudinal resistance $R_L$ (at $B_{ext} = 0T$) and the effective Hall carrier density (sheet density, see Supplementary Section 3), $c_{eff} = 1/(e\alpha)$, as functions of $V_g$ from -10V to +10V. Here, $\alpha$ is the out-of-plane ordinary Hall slope. Remarkably, the overall carrier type can be tuned from $p$-type to $n$-type, with $c_{eff}$ changing from $6.2 \times 10^{12} \text{ cm}^{-2}$ to $-3.5 \times 10^{12} \text{ cm}^{-2}$ ($-$ sign means $n$-type) when $V_g$ scans from -10V to +10V, and near $V_g = +3V$, $R_L$ develops a peak while $c_{eff}$ diverges. In the following, we analyze carefully the carrier distributions and transport details in the Cr-TI film under different gate voltages.

First, the ambipolar field effect (Fig. 2b) and $R_L$ versus temperature behavior (see Supplementary Section 1) observed in our Cr-TI film suggest that the Fermi level is located inside the bulk band gap. However, due to the Cr-doping [37,38], along with the presence of defects and random potential fluctuations [43] in the film, there are a considerable amount of hole carriers in the bulk, which makes the transport analysis quite difficult. Nevertheless, from the gate-controlled two surface Shubnikov-de Haas quantum oscillations observed in a non-doped control sample (see Supplementary Section 3), we learn that the top gate can majorly control the top surface carrier density in the Cr-TI film [37,38,44], while the bottom surface carrier density remains almost unchanged through the whole voltage range with the Fermi level located in the topological SS conduction band [44]. For convenience, in Fig. 2b we divide $V_g$ from -10V to +10V into three regions: I, II and III. In region I, the top surface is biased to the *p*-type regime with $c_{eff}$ changes by a rate of $9 \times 10^{10} cm^{-2}/V$ and meanwhile $R_L$ changes monotonically. The top surface accumulated hole mobility is $\mu_h = 85$ cm$^2$/(V S) (see Supplementary Section 3), and since the topological SS Dirac point is located close to the bulk valence (BV) band edge [44-46] in the band structure, a large amount of these holes are ordinary. In region III, the top surface is biased to the topological SS *n*-type regime and since $c_{eff}$ almost reaches the linear region, the overall topological SS electrons (from both top and bottom surfaces) dominate the transport. Considering both $c_{eff}$ and $R_L$, the topological SS Dirac electron mobility can be estimated as $\mu_D = 100$ cm$^2$/(V S). In the intriguing region II, $R_L$ reaches a peak at around $V_g = +3V$ which means the top surface carriers are mostly depleted, and at the same time $c_{eff}$ diverges. Solving the equations set by the divergence of $c_{eff}$ and the maximum of $R_L$ (see Supplementary Section 3), we find that the bottom topological SS electron density is $n_{SS,b} = 1.02 \times 10^{12} cm^{-2}$ and the bulk hole density (sheet density) is $p_{bulk} = 1.4 \times 10^{12}$ cm$^{-2}$ at $V_g = +3V$.

Due to the randomness of defects and potential fluctuations in the Cr-TI film [43], the bulk carriers are assumed to be uniformly distributed inside the bulk and consequently cannot produce much SOT when passing a charge current through the film. As a result, we will focus mainly on the carriers at the two surfaces which are the dominant origins of the current-induced SOT. In Fig. 2c, we depict both the top and bottom surface band structures for the corresponding three regions: I, II and III. While the bottom surface Fermi level $E_F$ always remains inside the topological SS conduction band, the top surface Fermi level $E_F$ can be tuned near the bulk VB edge (region I), across the surface gap near the Dirac point (region II), and into the topological SS conduction band (region III). In region III, even though the top surface Fermi level $E_F$ is tuned into the topological SS conduction band, the top SS electron density $n_{SS,t}$ is still estimated to be smaller than the bottom SS electron density $n_{SS,b}$ (more evidences can be found in the Discussions section).

In addition to the manipulation of surface carrier density and type, the out-of-plane anisotropy coefficient $K$ and the saturation AHE resistance $R_A$ of the Cr-TI film can also be significantly modified by the gate voltage. Figure 2d shows $K$ and $R_A$ as functions of the gate voltage $V_g$ from -10V to +10V. It should be noted that the $R_A$ versus $V_g$ curve shows a similar shape to the $R_L$ versus $V_g$ curve as displayed in Fig. 2b, which is in accordance with the intrinsic correlation between $R_A$ and $R_L$ due to different scattering mechanisms [47]. Equally important, the anisotropy coefficient $K$ shows almost a monotonic decrease as $V_g$ tunes the top surface carriers from holes to electrons, and the $K$ value is modulated from 719.5 mT ($V_g = -10$V) to 471.2 mT ($V_g = +6$V), which has changed by 35% and suggests that electric field can have a crucial modulation of the magnetism in the Cr-TI film [37,38,42,48].

**Electric-field control of SOT in the Au(electrode)/Al$_2$O$_3$/Cr-TI/GaAs(substrate) structure.** Based on the above measured parameters, we carried out the second harmonic experiment to probe the effective spin-orbit field $B_{SO}$ under different gate voltages. In Fig. 3a, we present the second harmonic AHE resistance $R_{AHE}^{2\omega}$ as a function of the in-plane magnetic field for different gate voltages from -10V to +10V. The AC current applied is 4μA (*rms* value) and the low frequency used is 15.8Hz. The solid lines represent the fittings proportional to $1/(|B_y| - K)$ in the large field region according to equation (1). In Fig. 3b, we plot the obtained effective spin-orbit field $B_{SO}$ versus the applied gate voltage for the $B_y > 0$ case. Remarkably, the effective spin-orbit field $B_{SO}$ can be tuned from 19.8 mT ($V_g = -10V$) up to 79.5 mT ($V_g = +3V$), reflecting a factor of 4 in modulation which is almost two orders of magnitude larger than that reported in the gate-controlled HMFHs [10]. $B_{SO}$ is peaked at around $V_g = +3V$, which corresponds to the situation when the top surface of the Cr-TI film is most insulating (neutral point) and the spin-polarized current mainly flows through the bottom topological SS [24-28] and consequently produces the largest SOT. When $V_g > +3V$, e.g., the Cr-TI film enters region III as shown in Fig. 2b, the top surface can also have spin-momentum locked SS electrons which will share the current flowing through the film. However, these top SS electrons will generate opposite SOT compared to the bottom SS electrons due to the different spin-momentum locking direction with respect to the bottom surface normal vector [24-28]. On the other hand, when $V_g < +3V$, e.g., the Cr-TI film enters region I as shown in Fig. 2b, the accumulated holes on the top surface will also share the current flowing through the film, and a large amount of these holes are ordinary and hence cannot generate much SOT. The rest amount of holes are Dirac-like and similarly will generate opposite SOT compared to the bottom SS electrons. In both cases ($V_g$ gets larger or smaller from +3V), the overall current-induced SOT in the Cr-TI film decreases as evidenced in Fig. 3b. It is noted that the overall current-induced SOT does not change sign in the whole region from -10V to +10V, indicating that the bottom SS electrons always dominate in generating the SOT. Combined with the modulation of magnetization magnitude $M_S$ by the gate voltage (Supplementary Section 6), the SOT strength can vary from $7.4 \times 10^{11}(\frac{\hbar}{2e})$ A cm$^{-3}$ (at $V_g = -10V$) to $2.87 \times 10^{12}(\frac{\hbar}{2e})$ A cm$^{-3}$ (at $V_g = +3V$) within the accessible voltage range, which demonstrates the effective gate electric-field control of SOT (by a factor of 4) in the Au(electrode)/Al$_2$O$_3$/Cr-TI/GaAs structure.

For the purpose of potential applications, we further investigated the gate electric-field effect on the magnetization switching behaviors of the Cr-TI film in the presence of both the in-plane magnetic field $B_y$ and the longitudinal DC current $I_{dc}$. Intriguingly, the switching behaviors can be significantly modified by the gate voltage in both the ($I_{dc}$-fixed, $B_y$-dependent) and the ($B_y$-fixed, $I_{dc}$-dependent) magnetization switching experiments. For example, in Fig. 3c, we summarize the switching phase diagrams under $V_g = -10V, +1.5V, +10V$, respectively. It can be seen that the switching phase diagram has been changed dramatically by the gate voltage and the switching boundaries shrink towards the central region for $V_g = +1.5V$ compared with the other two cases ($V_g = -10V, +10V$), indicating a smaller longitudinal $I_{dc}$ is required for switching for $V_g = +1.5V$ in the presence of a fixed in-plane $B_y$. In fact, this is expected because the current-induced SOT efficiency is higher for $V_g = +1.5V$ than the other two cases ($V_g = -10V, +10V$). Furthermore, we carried out the ($I_{dc}$-fixed, $B_y$-fixed, $V_g$-dependent) experiments to study the direct manipulation of magnetization by scanning the gate voltage $V_g$. The initial magnetization state can be prepared to be $M_Z < 0$ for X: ($I_{dc} = 20\mu A$, $B_y = 0.1T$) and $M_Z > 0$ for Y: ($I_{dc} = 20\mu A$, $B_y = -0.1T$) under $V_g = -10V$, as shown in Fig. 3c. After that, both $I_{dc}$ and $B_y$ are fixed and we scan $V_g$ from -10V to +3V while the AHE resistance $R_{AHE}$ is simultaneously measured. The obtained results are plotted in Fig. 3d, and it can be clearly seen that the magnetization state is switched from $M_Z < 0$ to $M_Z > 0$ for X: ($I_{dc} = 20\mu A$, $B_y = 0.1T$) and from $M_Z > 0$ to $M_Z < 0$ for Y: ($I_{dc} = 20\mu A$, $B_y = -0.1T$) by scanning the gate voltage $V_g$ from -10V to +3V, successively. The final magnetization states are consistent with the deterministic states for X and Y under $V_g = +3V$, respectively, as shown in Fig. 3c. Consequently, the switching induced by scanning the gate voltage in the presence of fixed $I_{dc}$ and $B_y$ demonstrates that the gate electric-field control of SOT can provide an effective way to determine and manipulate the magnetization state in the Cr-TI film.

**Discussions.** Now, we attempt to explore the correlations between the surface carrier densities, surface currents, surface band structures and the measured electric-field control of SOT in the top-gate Hall bar device. As discussed before, the bottom topological SS electron density $n_{SS,b}$ remains unchanged through the whole voltage range while the top surface carriers can be tuned from *p*-type to *n*-type by $V_g$. Based on a simple capacitor model and the carrier-attracting rate by the top gate, we plot both the top and bottom surface carrier densities as functions of $V_g$ in Fig.4 upper panel. From -10V to +2.3V, the top surface accumulated hole density $p_{\text{top}}$ drops linearly and from +3.3V to +10V, the top topological SS electron density $n_{\text{SS,t}}$ increases linearly. The region $+2.3\text{V} < V_g < +3.3\text{V}$ corresponds to the plateau window in $R_L$ in the same voltage region as shown in Fig. 2b. In Fig.4 upper panel, we also depict the corresponding top and bottom surface band structures for different voltage regions. It is clear that $n_{\text{SS,b}}$ is always larger than $n_{\text{SS,t}}$ in the whole *n*-type regime. Since the top and bottom topological SS carriers have opposite spin-momentum locking directions, we define a net spin-polarized surface current $I_{\text{SS}}^{\text{net}} = I_{\text{SS}}^{\text{bot}} - I_{\text{SS}}^{\text{top}}$, where $I_{\text{SS}}^{\text{top}}$ and $I_{\text{SS}}^{\text{bot}}$ are the top and bottom SS currents, respectively. Then, the ratio $I_{\text{SS}}^{\text{net}}/I_{\text{tot}}$ ($I_{\text{tot}}$ is the total current passing through the film) will quantify the percentage of current that is responsible for producing the SOT. In the region $+3\text{V} < V_g < +10\text{V}$, $\frac{I_{\text{SS}}^{\text{net}}}{I_{\text{tot}}} = (n_{\text{SS,b}} - n_{\text{SS,t}})e\mu_D R_S$, where $R_S$ is the sheet resistance of the film. In Fig. 4 lower panel, we plot both the $I_{\text{SS}}^{\text{net}}/I_{\text{tot}}$ ratio and the current-induced effective spin-orbit field $B_{\text{SO}}$ which is obtained from Fig. 3b. Surprisingly, the $I_{\text{SS}}^{\text{net}}/I_{\text{tot}}$ ratio and the measured $B_{\text{SO}}$ show a quite similar trend in $+3\text{V} < V_g < +10\text{V}$ with $I_{\text{SS}}^{\text{net}}/I_{\text{tot}}$ changing from 0.46 to 0.12 and $B_{\text{SO}}$ changing from 79.5mT to 22mT, indicating that $I_{\text{SS}}^{\text{net}}$ is indeed the origin of the induced SOT. In the $-10\text{V} < V_g < +3\text{V}$ region, by assuming that 57% of the top surface accumulated holes are Dirac holes, the $I_{\text{SS}}^{\text{net}}/I_{\text{tot}}$ ratio can also exhibit a trend similar to the $B_{\text{SO}}$ curve. Now it is time to derive the intrinsic spin-torque ratio from the net spin-polarized surface current: $\vartheta_{\text{ST}} = \frac{2eM_S B_{\text{SO}} w t_S t_{\text{Cr-TI}}}{\hbar (I_{\text{SS}}^{\text{net}}/I_{\text{tot}}) I_{\text{ac}}^{\text{peak}}}$, where $t_S \cong 1.5\text{nm}$ is the surface states penetration depth [49]. Taking in all the parameters, we find $\vartheta_{\text{ST}} = 116$ and it is independent of $V_g$. The large $\vartheta_{\text{ST}}$ value again shows the giant spin-torque efficiency that the TI topological SS possess over conventional materials, such as HMFHs.

In conclusion, we have demonstrated the current-induced giant SOT and magnetization switching in the uniformly doped Al$_2$O$_3$/Cr-TI/GaAs structure, which is due to the structural asymmetry at the two interfaces, and the effective electric-field control of SOT in the Au(electrode)/Al$_2$O$_3$/Cr-TI/GaAs structure by manipulating the top surface carrier density and type through electrical gating. Through detailed analysis, we find that the current-induced SOT has a strong correlation with the net spin-polarized surface current originating from the topological SS in the Cr-TI film. The large gate tuning of SOT (a factor of 4) and the gate voltage-induced magnetization switching in the presence of constant in-plane $B_y$ and $I_{dc}$ suggest that Cr-TI materials may find wide implications in the gate-controlled spin-torque devices that are compatible with modern field-effect semiconductor technologies. Here, we would like to mention that by engineering the interface roughness of the film, the current-induced SOT can also be dramatically changed, which again demonstrates the interfacial SS origin of the SOT (for details, see Supplementary Section 2). The effective electric-field control of SOT, together with the giant spin-torque efficiency of TI [15-17] and the high quality of TI/magnetic insulator structures [50], suggests that TI-based magnetic structures may potentially lead to the development of novel spintronic devices which exhibit ultralow power consumption and new functionalities such as simultaneous memory and logic functions.

**Methods**

**Material growth.** The uniformly Cr-doped Cr$_{0.16}$(Bi$_{0.50}$Sb$_{0.42}$)$_2$Te$_3$ thin film was grown in an ultrahigh vacuum Perkin Elmer MBE system. Semi-insulating ($\rho > 10^6$ Ω·cm) GaAs (111)B substrate has been cleaned by acetone with ultrasonic for 10 minutes before loaded to the MBE chamber. Then the atomically flat substrate was annealed to 580°C and cooled down to growth temperature, under Se rich environment. High-purity Bi (99.9999%), Te (99.9999%), Cr (99.99%) and Sb (99.999%) were evaporated by conventional effusion cells and cracker cells. During the Cr-TI film growth, the GaAs (111)B substrate was maintained at 200°C (growth temperature). Bi and Te cells were kept at 470°C and 320°C, respectively, and the Sb and Cr temperatures were carefully chosen in order to control the carrier density and the magnetic doping profile in the film. We used an *in-situ* real-time high-energy electron diffraction (RHEED) technique to monitor the epitaxial growth. The film surface was found to be atomically flat as evidenced by the streaky RHEED patterns. After the Cr-TI film growth, a 1nm thick $Al_2O_3$ capping layer was immediately grown *in-situ* on top to protect the film for following fabrications.

**Characterizations.** (1) **Energy-dispersive X-ray (EDX) mapping.** The EDX elemental mapping image was obtained by an EDX spectroscope which is attached to a high-resolution FEI TITAN Cs-corrected scanning transmission electron microscopy (STEM) operating at 200 KV. (2) **Transport measurements.** Four-point Hall measurements were performed using the Quantum Design physical property measurement system (PPMS). Several experimental parameters such as temperature, measurement frequency, magnetic field, and external gate voltage can be systematically altered. Multiple lock-in-amplifiers and Keithley source meters were connected with the PPMS system, leading to high-sensitivity transport measurements for all the Hall bar devices.

**Device fabrication.** The uniformly Cr-doped TI thin film was patterned into micron-scale Hall bar devices using conventional photolithography with subsequent $CHF_3$ dry etching for 18 s. A 20 nm thick high-*k* $Al_2O_3$ dielectric layer was deposited by atomic layer deposition (ALD) at 200°C. For the Hall bar device without the top gate, the Hall channel contacts were defined by *e*-beam evaporation after the $Al_2O_3$ was etched away in the contact areas. A layer of Au (100 nm) was directly deposited onto the exposed TI top surface to form the contacts. For the Hall bar device with the top gate electrode, a top-gate metal scheme of Au (100 nm) was achieved with a second step of photolithography and *e*-beam evaporation.

**Acknowledgements**

The material growth and characterizations were supported by the DARPA Meso program under contract No.N66001-12-1-4034 and N66001-11-1-4105. The device fabrication and low temperature measurements were supported as part of the Spins and Heat in Nanoscale Electronic Systems (SHINES), an Energy Frontier Research Center funded by the U.S. Department of Energy (DOE), Office of Science, Basic Energy Sciences (BES), under Award # DE-SC0012670. The analysis and theoretical modeling were supported by the U.S. Army Research Office under grants W911NF-14-1-0607 and W911NF-15-1-0561. We are also very grateful to the support from the FAME Center, one of six centers of STARnet, a Semiconductor Research Corporation program sponsored by MARCO and DARPA. Y.W. thanks the support of the National 973 Program of China (2013CB934600), National Science Foundation of China (11174244, 51390474) and Zhejiang Provincial Natural Science Foundation of China (LR12A04002).

**Figure Legend**

**Figure 1 | Current-induced magnetization switching and second harmonic measurements in the Al$_2$O$_3$(20nm)/Cr-TI(7nm)/GaAs(substrate) structure device. a**, 3D schematic of the Al$_2$O$_3$/Cr-TI/GaAs structure. Inside the Cr-TI layer, the blue arrows denote the Cr dopants and the red arrows indicate the spin directions of the conducting topological SS carriers at the two interfaces when passing a negative DC current (*i.e.,* along $-y$ direction) through the film. Shown on top of the structure are the torques exerted by the external magnetic field $\boldsymbol{B}_{\text{ext}}$, the anisotropy field $\boldsymbol{B}_{\text{K}}$, and the net current-induced SOT $\boldsymbol{\tau}_{\text{SO}}$. $\boldsymbol{B}_{\text{ext}}$ and magnetization $\boldsymbol{M}$ are both in the $yz$ plane. Right: the top surface band structure and the bottom surface band structure. "BC", "BV" and "SS" stand for bulk conduction band, bulk valence band and surface states, respectively. $E_{\text{F}}$ is the Fermi level. **b**, Transverse Hall resistance $R_{\text{H}}$ and longitudinal resistance $R_{\text{L}}$ as functions of the out-of-plane magnetic field at 1.9K. Upper inset: configuration of the applied magnetic field. Lower inset: microscopic image of the Hall bar device with illustrations of the Hall measurement set-up: current flowing from the left to the right (along $y$-direction) is defined as the positive current; $V_{\text{H}}$ measures the Hall voltage and $V_{\text{L}}$ measures the longitudinal voltage. The width of the Hall bar and the length between two neighboring Hall contacts are both 10 μm. **c**, ($B_y$ -fixed, $I_{\text{dc}}$ -dependent) magnetization switching experiment. The current-induced magnetization switching is measured in the presence of a constant in-plane magnetic field with $B_y = +2$ T and $B_y = -2$ T, respectively. **d**, Switching phase diagram of the magnetization in

the presence of both $B_y$ and $I_{dc}$. The dashed lines and symbols (extracted from experiments) denote the boundaries between different states. **e**, Second harmonic AHE resistance as a function of the in-plane magnetic field for AC current with different *rms* values. The frequency used is 15.8 Hz. Solid lines indicate the fittings proportional to $1/(|B_y| - K)$ in the large field regions. **f**, Effective spin-orbit field $B_{SO}$ as a function of the AC current $I_{ac}$ (*rms* value) for both the $B_y > 0$ and $B_y < 0$ cases as extracted from **e**. Straight lines are the linear fittings. Error bars represent standard errors. All the measurements are performed at 1.9K.

**Figure 2 | Top-gate Hall bar configuration and gate electric-field effect on material properties in the Au(electrode)/Al$_2$O$_3$(20nm)/Cr-TI(7nm)/GaAs(substrate) structure device.** **a**, 3D schematic illustration of the Hall bar structure made from the Al$_2$O$_3$(20nm)/Cr-TI(7nm)/GaAs(substrate) stack with a top Au gate electrode (light gray). Standard four-point measurement setup is displayed. A gate voltage of $V_g$ can be applied between the top gate and the source contact. **b**, Longitudinal resistance $R_L$ and effective Hall carrier density (sheet density) as functions of $V_g$. Colored area I, II and III show the different gate voltage regions. $p$, $n_{SS}$ and $n_{SS,b}$ represent the overall hole density, the overall SS electron density and the bottom SS electron density in the film, respectively. **c**, The top surface and bottom surface band structure configurations for the three gate voltage regions shown in **b**. Yellow colored area in the top surface band structure shows the tune range of the top surface Fermi level $E_F$ within the corresponding gate voltage region. **d**, Out-of-plane anisotropy coefficient $K$ and out-of-plane saturation AHE resistance $R_A$ as functions of the gate voltage. Error bars represent standard errors. All the measurements are performed at 1.9K.

**Figure 3 | Second harmonic measurements under different gate voltages and voltage-induced magnetization switching behaviors.** **a**, Second harmonic AHE resistance as a

function of the in-plane magnetic field under different gate voltages from -10V to +10V. The AC current applied is 4µA (*rms* value) and the frequency used is 15.8 Hz. Solid lines represent the fittings proportional to $1/(|B_y| - K)$ in the large field regions. **b**, Effective spin-orbit field $B_{SO}$ as a function of the gate voltage for the $B_y > 0$ case as extracted from figure **a**. Error bars represent standard errors and blue curve shows the Lorentz fitting. Insets show the schemes of surface carrier distribution in the Cr-TI layer under $V_g = -10V, +3V$ and $+10V$. **c**, Magnetization switching phase diagrams under $V_g = -10V, +1.5V$ and $+10V$ in the presence of both $B_y$ and $I_{dc}$. The dashed lines and symbols (extracted from experiments) represent the boundaries between the different states. **d**, Magnetization switching induced by scanning gate voltage $V_g$ in the presence of constant $B_y$ and $I_{dc}$ for X: ($I_{dc} = 20\mu A$, $B_y = 0.1T$) and Y: ($I_{dc} = 20\mu A$, $B_y = -0.1T$). Insets show the corresponding initial and final magnetization configurations. All the measurements are performed at 1.9K.

**Figure 4 | Correlations between the surface carrier densities, surface currents, surface band structures and the measured electric-field control of SOT in the top-gate Hall bar device.** Upper panel: Top and bottom surface carrier densities as functions of $V_g$. Insets show the top and bottom surface band structures in different gate voltage regions. Yellow colored area in the top surface band structure shows the tune range of the top surface Fermi level $E_F$ within the corresponding gate voltage region. Lower panel: comparison between the ratio $I_{SS}^{net}/I_{tot}$ (net spin-polarized surface current over total current) and the effective spin-orbit field $B_{SO}$, as functions of gate voltage $V_g$. The net spin-polarized surface current is defined as $I_{SS}^{net} = I_{SS}^{bot} - I_{SS}^{top}$. $B_{SO}$ is re-plotted here as from Fig. 3b. Error bars represent standard errors and blue curve shows the Lorentz fitting.

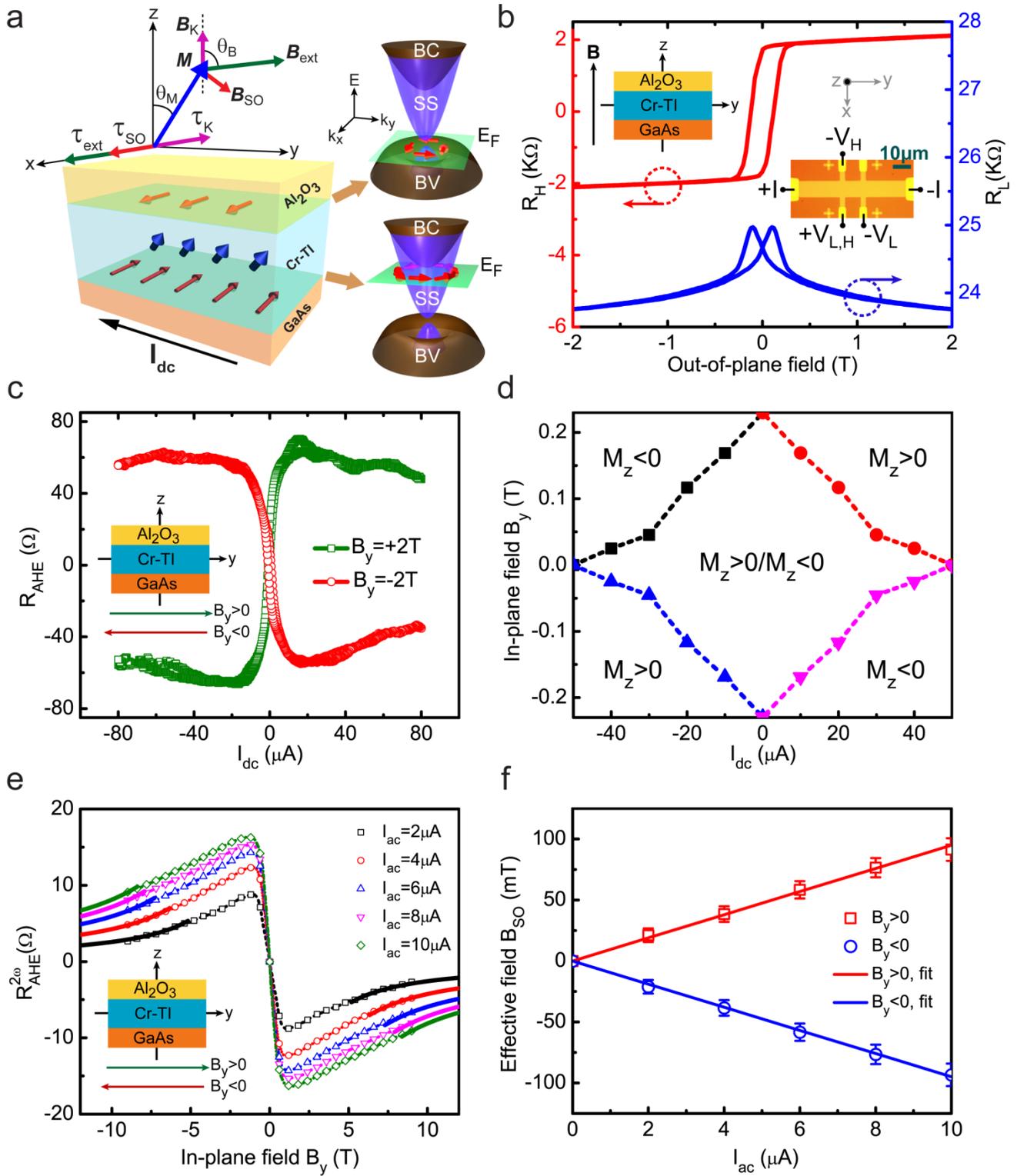

**Figure 1. by Yabin Fan *et al.***

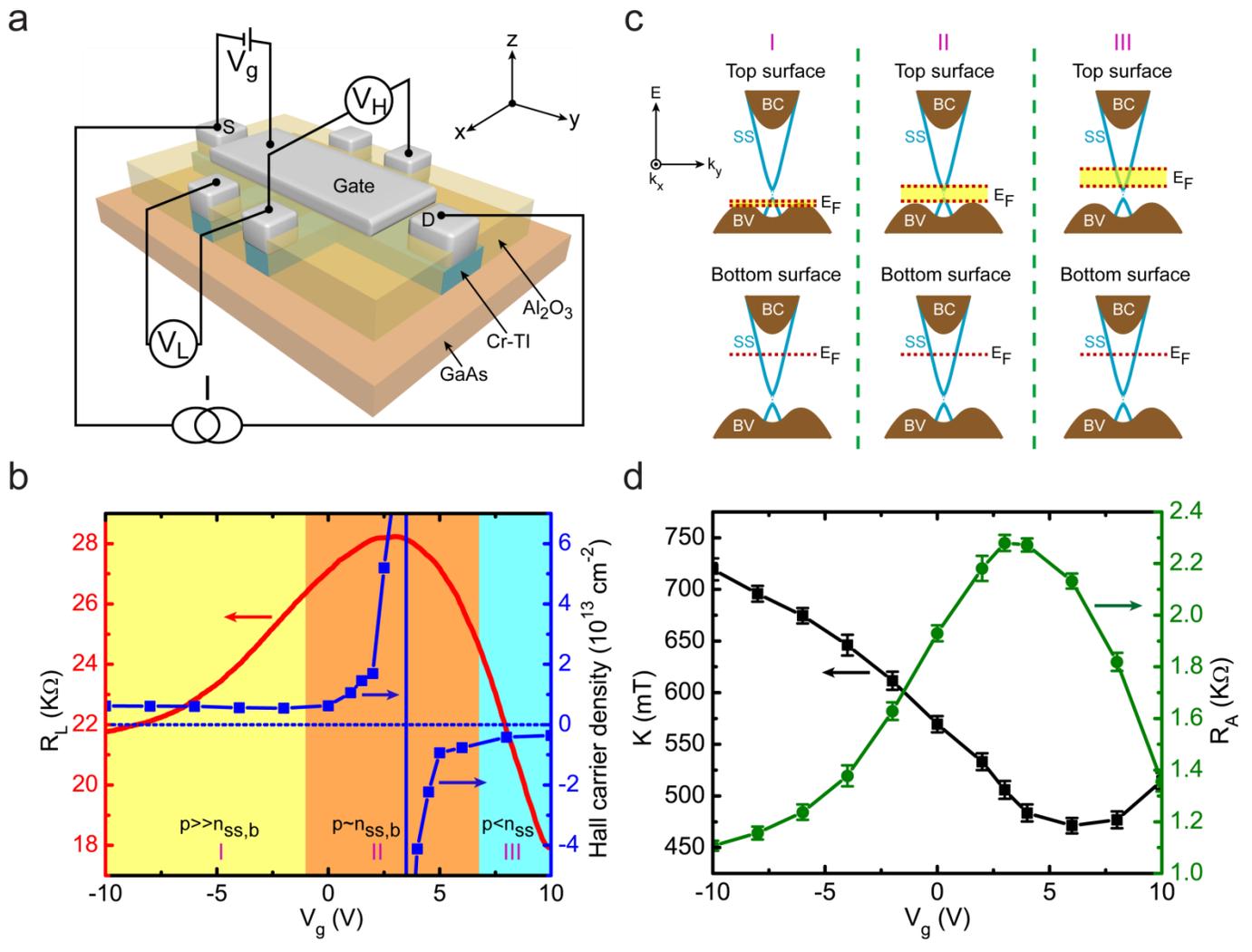

**Figure 2. by Yabin Fan *et al*.**

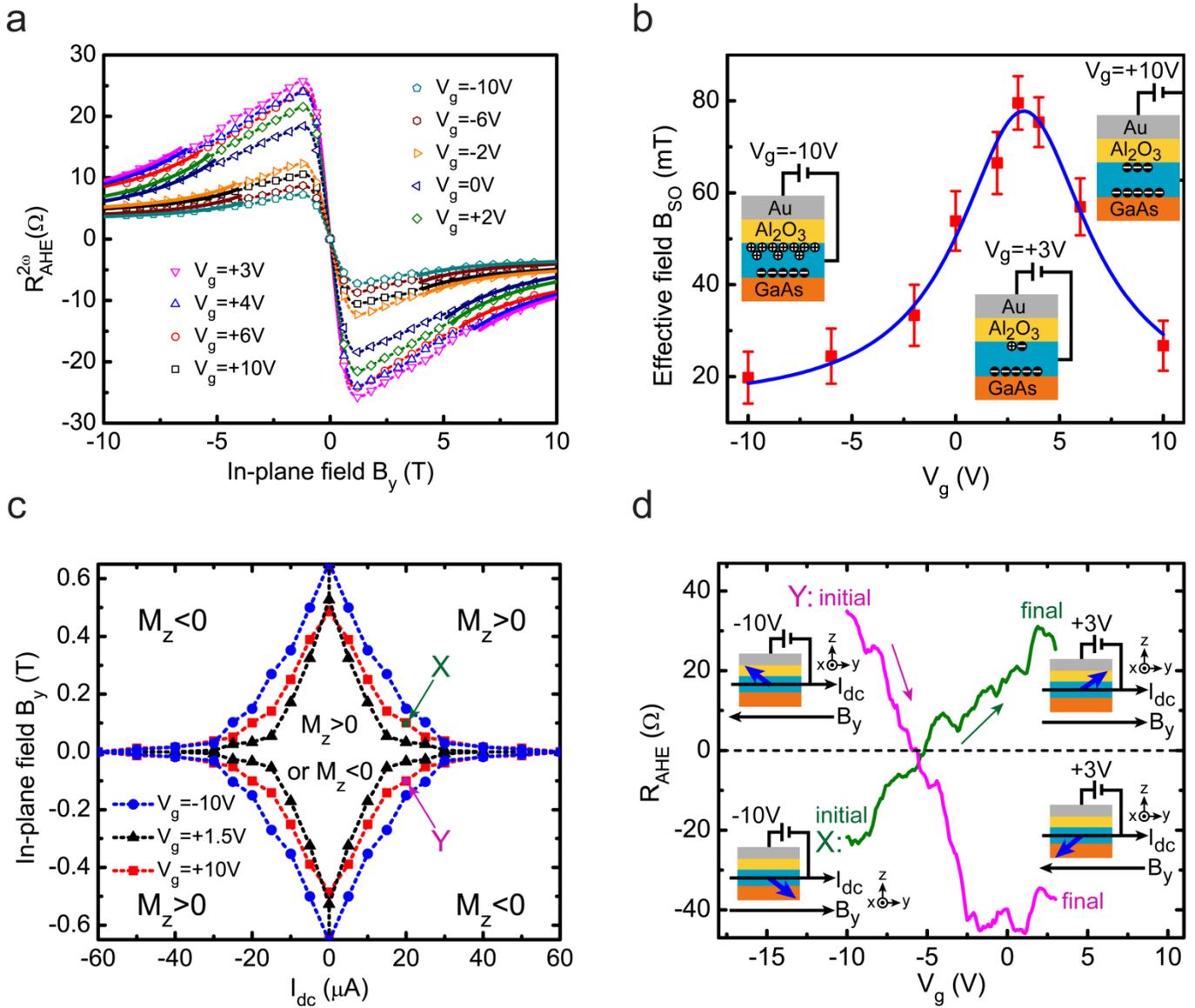

**Figure 3. by Yabin Fan *et al*.**

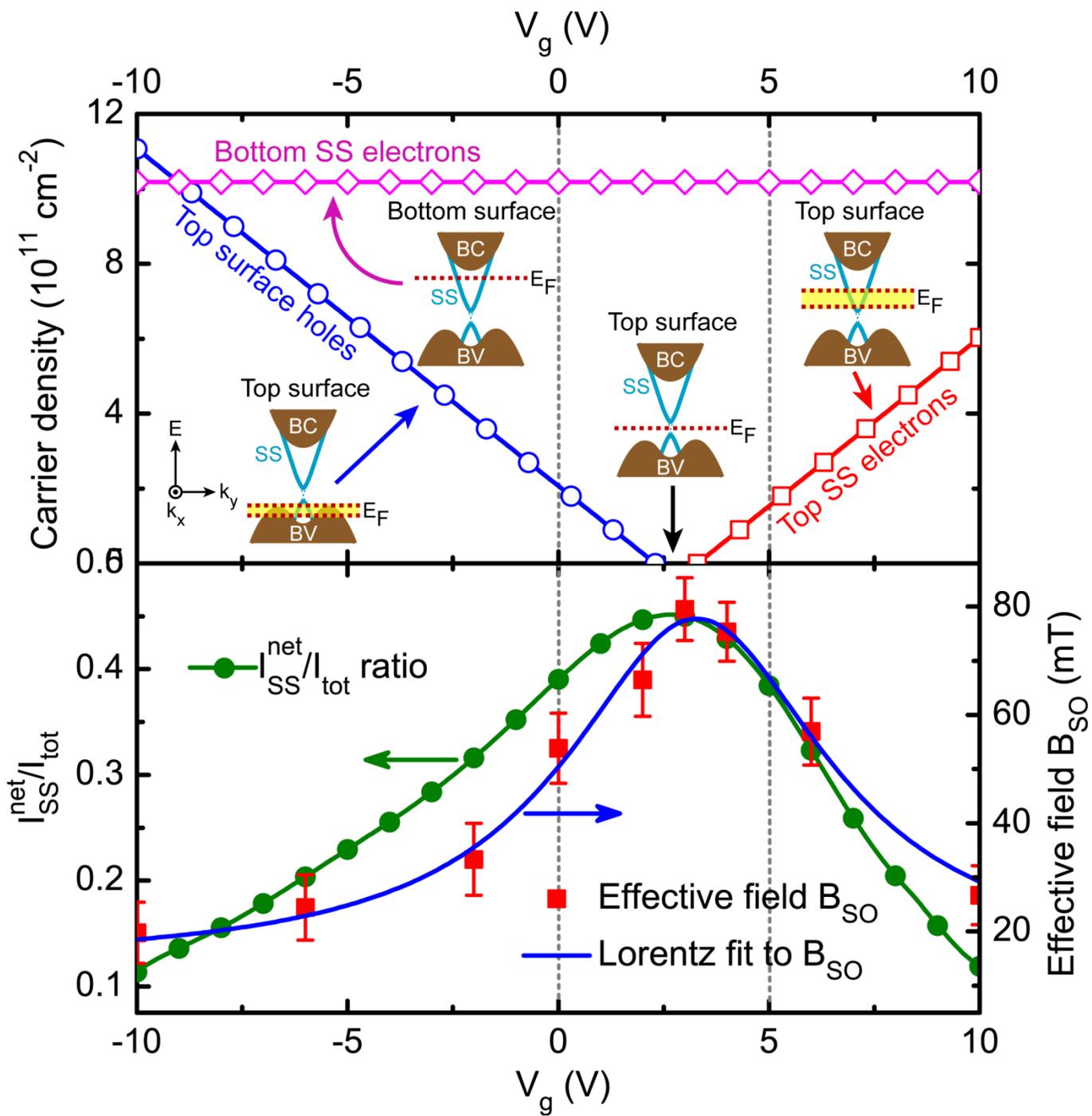

**Figure 4. by Yabin Fan *et al*.**

# Electric-field control of spin-orbit torque in a magnetically doped topological insulator

Yabin Fan[1†*], Xufeng Kou[1†], Pramey Upadhyaya[1†], Qiming Shao[1], Lei Pan[1], Murong Lang[1], Xiaoyu Che[1], Jianshi Tang[1], Mohammad Montazeri[1], Koichi Murata[1], Li-Te Chang[1], Mustafa Akyol[1], Guoqiang Yu[1], Tianxiao Nie[1], Kin L. Wong[1], Jun Liu[3], Yong Wang[3], Yaroslav Tserkovnyak[2] and Kang L. Wang[1*]

[1]Department of Electrical Engineering, University of California, Los Angeles, California 90095, USA

[2]Department of Physics and Astronomy, University of California, Los Angeles, California 90095, USA

[3]Center of Electron Microscopy and State Key Laboratory of Silicon Materials, School of Materials Science and Engineering, Zhejiang University, Hangzhou 310027, China

[†]These authors contributed equally to this work.

*To whom correspondence should be addressed. E-mail: yabin@seas.ucla.edu; wang@seas.ucla.edu

## SUPPLEMENTARY INFORMATION

**Contents:**

## 1. Material properties and measurement set-up

In this section, we provide some additional information on the material properties and measurement set-up in the Al$_2$O$_3$/Cr-TI/GaAs(substrate) structure, which will supplement the first part of the main text. Figure S1a shows the 3-dimensional (3D) schematic of the Al$_2$O$_3$(20nm)/Cr-TI(7nm)/GaAs(substrate) structure. The blue arrows indicate the Cr dopant elements inside the TI matrix and the right panel of Fig. S1a shows the scanning transmission electron microscopy (STEM) image of the Cr-TI film and the energy-dispersive X-ray (EDX) mapping of the Cr dopant elements in the Cr-TI layer which are uniformly distributed [1,2]. The STEM image demonstrates the nice crystallinity of the film and the atomically sharp interface between the Cr-TI layer and the substrate. The film was patterned into micron-scale Hall bar structures and Fig. S1b shows the detailed microscopic image of the Hall bar device with illustrations of the Hall measurement set-up.

To examine the quality of the Cr-TI thin film, temperature-dependent longitudinal resistance $R_L$ at zero magnetic field was measured and it shows monotonic decrease when increasing temperature from 1.9K to 300K, as plotted in Fig. S1c, which indicates the semiconducting feature of the film and suggests the Fermi level is located inside the bulk band gap [3]. Also shown in Fig. S1c is the coercive field $B_C$ versus temperature, from which the Curie temperature $T_C$ of the film is estimated to be around 11K where $B_C$ almost vanishes.

In order to demonstrate the current-induced spin-orbit torque (SOT) arising from the non-balanced two interfaces in the Hall bar structure, we also carried out the ($I_{dc}$-fixed, $B_y$-dependent) magnetization switching experiments. As shown in Fig. S1d, the out-of-plane component of the magnetization, $M_Z$ (manifested by the anomalous Hall effect (AHE) resistance $R_{AHE}$), can be successfully switched by sweeping the in-plane magnetic field in the presence of the fixed DC currents $I_{dc} = \pm 20\mu A$. The switching is hysteretic and agrees with the definitions of $\boldsymbol{\tau}_{SO}$ and $\boldsymbol{B}_{SO}$ in Fig. 1a in the main text.

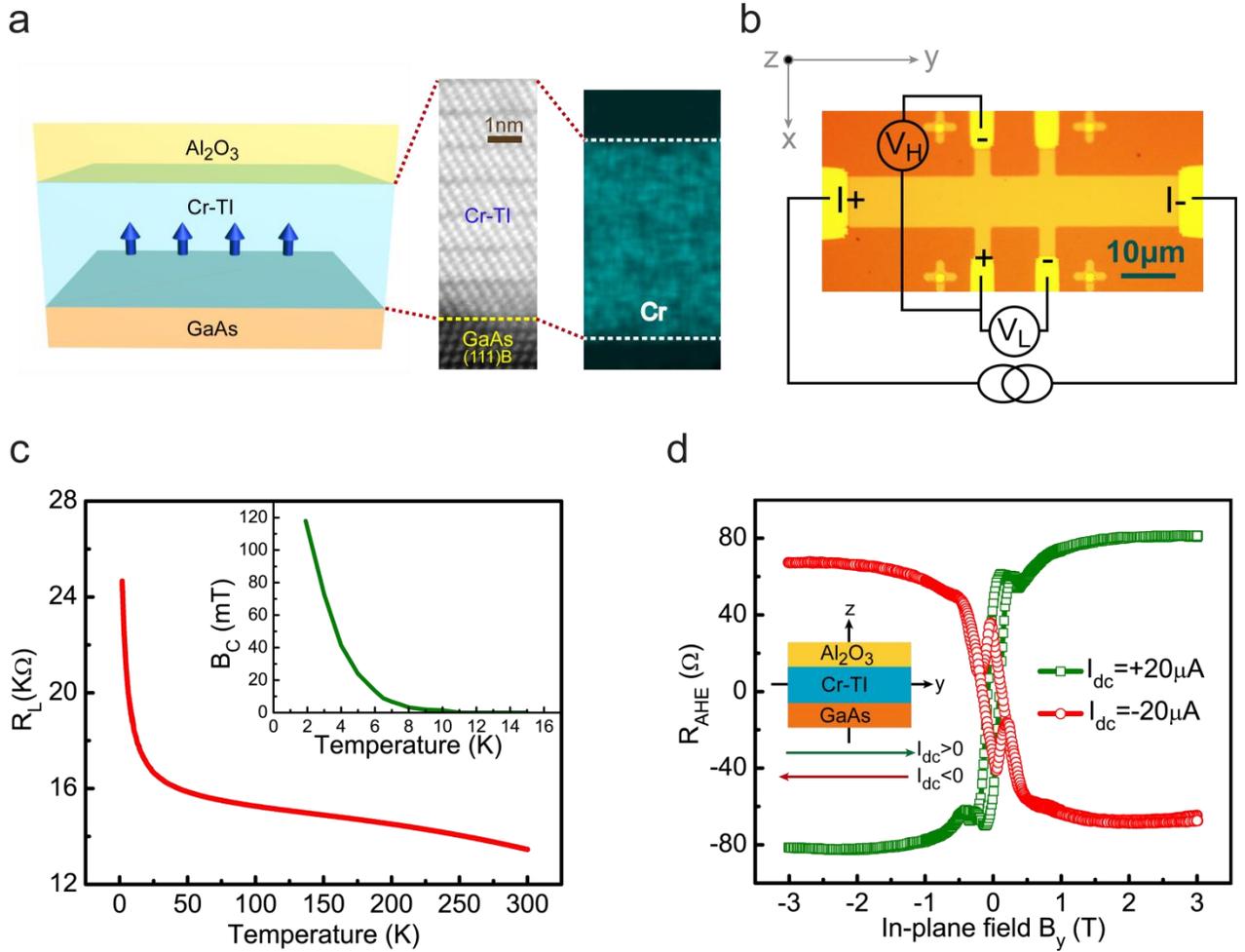

**Figure S1. Material properties and measurement set-up of the Al$_2$O$_3$(20nm)/Cr-TI(7nm)/GaAs(substrate) structure device. a**, 3D schematic of the Al$_2$O$_3$/Cr-TI/GaAs structure. The blue arrows denote the Cr dopants. Right: STEM image of the Cr-TI film and EDX mapping of the Cr dopant elements inside the Cr-TI layer. **b**, microscopic image of the Hall bar device with illustrations of the Hall measurement set-up: current flowing from the left to the right (along $y$-direction) is defined as the positive current; $V_H$ measures the Hall voltage and $V_L$ measures the longitudinal voltage. The width of the Hall bar and the length between two neighboring Hall contacts are both 10 μm. **c**, Longitudinal resistance $R_L$ versus temperature when the external magnetic field is set to 0T. Inset: the coercive field $B_C$ as a function of temperature. **d**, ($I_{dc}$-fixed, $B_y$-dependent) magnetization switching experiment. The AHE resistance is measured while sweeping $B_y$ in the presence of a constant DC current with $I_{dc} = +20$μA and $I_{dc} = -20$μA along the Hall bar, respectively.

## 2. Comparison of the current-induced spin-orbit torques in Cr-TI films grown on GaAs substrates with smooth and rough surface morphologies

In this section, we address the influence of interface qualities on the surface states (SS) transport and current-induced spin-orbit torque (SOT) in the Cr-TI film in the $Al_2O_3$/Cr-TI/GaAs(substrate) structure. In the main text, the atomically flat GaAs (111)B substrate was annealed at 580 °C for 5 minutes under Se rich environment (*i.e*, with Se vapor protection) to remove native surface oxide on the substrate. During this procedure, a strained GaSe single atomic layer was formed on the surface, which can protect the substrate surface to be flat after the pre-annealing and improve the growth of the Cr-TI film later on [4]. If we perform the pre-annealing without the Se vapor protection, due to the evaporation of As elements, the GaAs surface will evolve into a 3D feature surface structure after reconstruction. In Fig. S2a, we show the real-time high-energy electron diffraction (RHEED) patterns of the GaAs substrates after pre-annealing with and without Se vapor protection. It is obvious that the one with Se vapor protection shows streaky patterns which are the 2D feature of the substrate surface, suggesting the surface is atomically flat. In contrast, the one without Se vapor protection shows dotted patterns which are the 3D feature of the substrate surface. This indicates the pre-annealing without Se vapor protection can make the GaAs surface quite rough.

After the pre-annealing, a Cr-TI film was grown on the rough GaAs substrate following the same procedure as used in the main text. Here, we present both the STEM and atomic force microscope (AFM) studies of the Cr-TI films grown on the smooth GaAs substrate (as prepared in the main text) and on the rough GaAs substrate (pre-annealed without Se vapor protection). As can be seen in Fig. S2b, the Cr-TI grown on the smooth GaAs substrate shows very nice crystallinity and atomically sharp interface with the substrate. However, the Cr-TI grown on the rough GaAs substrate shows poor crystallinity and a lot of defects at the interface, as indicated by the rectangular box in Fig. S2b right panel. When preparing the samples for the STEM study, the top $Al_2O_3$ protection capping layer was removed and a Pt metal layer was deposited on top of the Cr-TI films for the Focused Ion Beam (FIB) process [2]. As a result, we cannot directly study the interface between the capping layer Al$_2$O$_3$ and the Cr-TI film by STEM. Alternatively, we performed the AFM studies of the two Cr-TI films (one grown on smooth GaAs substrate and one on rough GaAs substrate) to probe the top surface morphologies. We find the Cr-TI grown on the smooth GaAs substrate shows atomically flat surface morphology, as shown in Fig. S2c, while the Cr-TI grown on the rough GaAs substrate exhibits many pinholes, as illustrated in Fig. S2d, presumably because the GaAs substrate is too rough. In Fig. S2e and f, we present the surface roughness of the two films along the lines as indicated in Fig. S2c and d, respectively. Again, the Cr-TI grown on the smooth GaAs substrate shows very small roughness while the Cr-TI grown on the rough GaAs substrate shows pinholes with the depth even larger than the Cr-TI thickness. This indicates that these holes are deep into the GaAs substrate and agrees with the 3D feature RHEED pattern as shown in Fig. S2a lower panel.

Furthermore, we have performed both the magneto-transport and second harmonic measurements in the Al$_2$O$_3$/Cr-TI/(rough GaAs substrate) Hall bar structure device. The Cr-TI film again shows pronounced ferromagnetism (as shown in Fig. S3a) and second harmonic signals (as shown in Fig. S3b). Intriguingly, the second harmonic anomalous Hall effect (AHE) resistance, $R^{2\omega}_{\text{AHE}}$, as shown in Fig. S3b, changes polarity compared with the one measured in the Al$_2$O$_3$/Cr-TI/(smooth GaAs substrate) Hall bar structure (see Fig. 1e in the main text). This means the current-induced SOT has changed sign. In Fig. S3c, we plot the effective spin-orbit field $B_{\text{SO}}$, extracted from Fig. S3b, as a function of the AC current $I_{\text{ac}}$ (*rms* value), and find the effective field versus current ratio is $\frac{|B_{\text{SO}}|}{I^{\text{peak}}_{\text{ac}}} = 2.9$ mT/μA by linear fitting, which is much smaller than the one measured in the Al$_2$O$_3$/Cr-TI/(smooth GaAs substrate) structure (see main text). One possible explanation for the change in the sign of the current-induced SOT in the Al$_2$O$_3$/Cr-TI/(rough GaAs substrate) structure is that the large amount of defects at the Cr-TI/(rough GaAs substrate) interface induce many spin-dependent scatterings which can contaminate the bottom surface spin-polarized current, and leaving the top surface spin-polarized current dominant in generating the SOT.

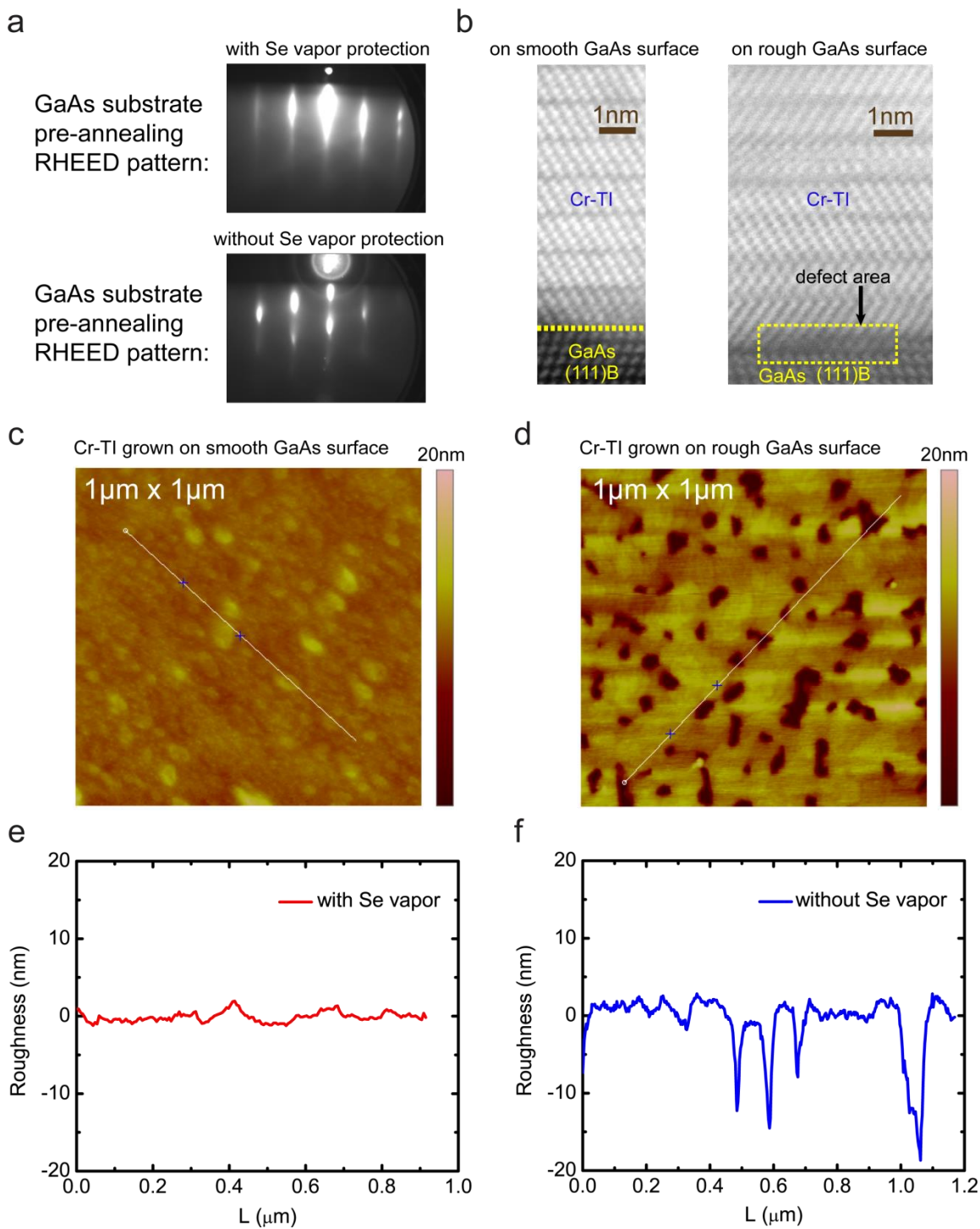

**Figure S2. Surface morphologies and interfacial properties of the Cr-TI films grown on GaAs (111)B substrates pre-annealed with / without Se vapor protection. a**, RHEED patterns of GaAs (111)B substrates after annealed at 580 ˚C for 5 minutes with / without Se vapor protection. **b**, STEM images of the Cr-TI films grown on smooth / rough GaAs substrates. **c** and **d**, AFM images of the Cr-TI films grown on smooth / rough GaAs substrates, respectively. **e** and **f**, Surface roughness of the Cr-TI films along the lines as indicated in **c** and **d**, respectively.

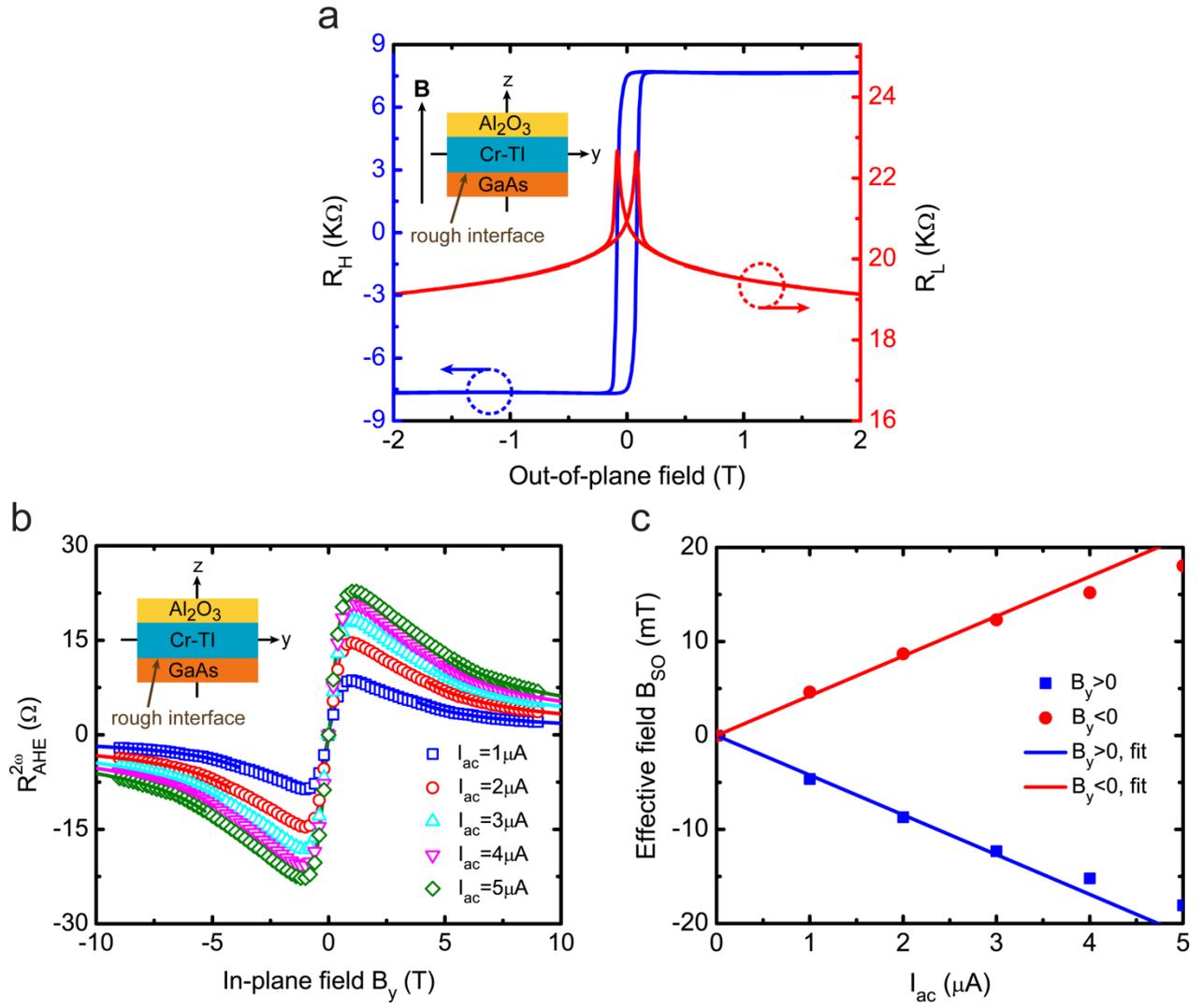

**Figure S3. Magneto-transport and second harmonic measurements in the Al$_2$O$_3$/Cr-TI/(rough GaAs substrate) structure. a**, Transverse Hall resistance $R_H$ and longitudinal resistance $R_L$ as functions of the out-of-plane magnetic field. Inset: configuration of the applied magnetic field. **b**, Second harmonic AHE resistance as a function of the in-plane magnetic field for AC current with different *rms* values. The frequency used is 15.8 Hz. Solid lines indicate the fittings proportional to $1/(|B_y| - K)$ in the large field regions. **c**, Effective spin-orbit field $B_{SO}$ as a function of the AC current $I_{ac}$ (*rms* value) for both the $B_y > 0$ and $B_y < 0$ cases as extracted from **b**. Straight lines are the linear fittings. All the measurements are performed at 1.9K.

## 3. Estimation of the top and bottom surface carrier densities and mobilities in the Cr-TI layer under different gate voltages

The best method to evaluate the top surface and the bottom surface carrier densities in the Cr-TI layer under different gate voltages is to study the quantum oscillations (*e.g.*, Shubnikov-de Haas (SdH) oscillations) [3-7] from these surface carriers when an out-of-plane magnetic field is applied. However, in the Cr-doped TI materials, since the carrier mobility is normally very low, as shown in Fig. S5, the relevant quantum oscillations are very difficult to observe. Alternatively, we have grown a 10-quintuple-layer (QL) un-doped $(Bi_{0.53}Sb_{0.47})_2Te_3$ thin film which has a similar Bi : Sb ratio to the Cr-TI film as used in the main text, and made it into a gate controllable Hall bar device following the same fabrication procedure. This un-doped TI thin film has a high carrier mobility (up to 3100 cm$^2$/(V·s) at 0.3K) [4], and the SdH oscillations from the two surface carriers can be successfully observed at low temperature. In Fig. S4a, we show the second derivative of the longitudinal resistance, $d^2R_L/dB^2$, as a function of the inverse of the magnetic field, $1/B$, and the gate voltage $V_g$ at 0.3 K. Intriguingly, we notice that there are $V_g$-dependent peaks as accentuated by the white dashed lines. These peaks originate from the formation of Landau levels of Dirac fermions on the top surface of the TI film, and the gate-dependent shift of the peaks is due to the modulation of the top surface carrier density as $V_g$ is scanned from +2V to +11V. At the same time, there are other $V_g$-independent peaks at high magnetic field, as indicated by the black dashed lines. These peaks are attributed to the formation of Landau levels of the bottom surface Dirac fermions. It is noted that the bottom surface SdH oscillation frequency remains almost constant within the whole gate voltage range ($-11V \leq V_g \leq +11V$), indicating little or almost no change of the bottom surface carrier density presumably due to the screening effect of the top surface carriers and the high dielectric constant ($\varepsilon_r \sim 75$) of TI materials [8,9].

The surface Dirac fermions feature can be confirmed from the Landau fan diagram (Fig. S4b) and the quadratic relationship between the Fermi level $E_F$ and the surface carrier density $n_{2D}$ (Fig. S4c). The Landau fan diagram for various gate voltage values is plotted in Fig. S4b, where the $1/B$ values corresponding to the peaks in Fig. S4a are plotted as a function of the Landau level index $n$ [3-7]. The solid symbols represent the top surface Dirac fermions, demonstrating a systematic shift depending on the gate bias. The open symbols represent the bottom surface Dirac fermions, which show almost no dependence on the gate bias. It is known that in the SdH oscillations, the Landau level index $n$ is related to the cross section area of the Fermi surface ($S_F$) by $2\pi(n+\gamma) = \hbar S_F/(eB)$, where $e$ is the electron charge, $\hbar$ is the reduced Plank constant, $B$ is the magnetic flux density, and $\gamma = 1/2$ or 0 represents the Berry phase of $\pi$ or 0 [3-7]. Linear fits yield intercepts at the abscissa of $0.51 \pm 0.04$, as shown in Fig. S4b, confirming the presence of surface massless Dirac fermions carrying a $\pi$ Berry phase. From the Landau fan diagram and the E-K dispersion relation from the ARPES results [4], the surface Fermi level $E_F$ and the surface carrier density $n_{2D}$ can be estimated. Figure S4c displays the Fermi level $E_F$ as a function of the surface carrier density $n_{2D}$. The $n_{top}$ can be effectively tuned from $1.1 \times 10^{12}$ cm$^{-2}$ to $1.6 \times 10^{12}$ cm$^{-2}$ by scanning the gate voltage from +2V to +11V, while the $n_{bottom}$ remains the same at $5.7 \times 10^{12}$ cm$^{-2}$ for all the gate voltages. A quadratic relationship of $E_F \propto n^{1/2}$ can be fitted, as shown by the red curve in Fig. S4c, confirming the linear E-K relationship of the surface Dirac cone.

In this non-doped TI Hall bar structure, the SdH oscillation frequency changes with respect to the gate voltage $V_g$, as shown in Fig. S4a, which demonstrates that the SdH oscillations are from the surface Dirac electrons (not holes). At $V_g = 0V$, the top surface SdH oscillation spectrum almost disappears, suggesting that the top surface Fermi level almost reaches the Dirac point at $V_g = 0V$. It is also found that at $V_g = 0V$ the whole film is at near the neutral state [4]. When $V_g < 0V$, as shown in Fig. S4a, the top surface SdH oscillation spectrum completely disappears while the bottom surface one still persists, indicating the accumulation of a large amount of ordinary holes on the top surface which reduce the overall carrier mobility [4]. When $0V < V_g < 11V$, the top surface has Dirac electrons but the density is always smaller than that of the bottom surface, as shown in Fig. S4c, which is probably due to the different interfacial properties (*e.g.*, different band bending) [10] at the two surfaces of the TI film.

Now we will estimate the carrier densities and mobilities on the top surface and bottom surface of the Cr-TI film in the Au(electrode)/Al$_2$O$_3$/Cr-TI/GaAs structure under different gate voltages. Compared with the un-doped TI film where the bulk is mostly insulating [4,8,9], the Cr dopants can induce hole carriers throughout the Cr-TI film and make the bulk non-insulating [1,2], as evidenced by the effective Hall carrier density $c_{\text{eff}}$ at $V_g = 0$V in Fig. 2b in the main text. Considering the value of $c_{\text{eff}}$ at $V_g = 0$V and the uniform Cr-doping profile in the Cr-TI film, it is reasonable to think that there are a considerable amount of topological SS electrons on the bottom surface of the Cr-TI film which are immune to the top gate voltage $V_g$, similar to the non-doped TI film as discussed above. In the main text, the effective Hall carrier density $c_{\text{eff}}$ (sheet density) is defined as, $c_{\text{eff}} = 1/(e\alpha)$, where $\alpha$ is the out-of-plane ordinary Hall slope. Generally, when there are both electrons and holes in a semiconducting film, the effective Hall carrier density can be derived as $c_{\text{eff}} = \frac{(p\mu_h + n\mu_e)^2}{p\mu_h^2 - n\mu_e^2}$, where $p$ is the hole density (sheet density), $n$ is the electron density (sheet density), $\mu_h$ is the hole mobility and $\mu_e$ is the electron mobility.

When the Cr-TI film is biased to the *p*-type regime, as shown in Fig. 2b region I in the main text, $c_{\text{eff}}$ changes almost linearly with $V_g$. In this regime, the top gate can most effectively control the top surface accumulated hole density because of the screening effect characterized by the small Debye length [1]. In this case, $\frac{dp_{\text{top}}}{dV_g} \cong \frac{dp}{dV_g} \cong \frac{dc_{\text{eff}}}{dV_g} = 9 \times 10^{10}$ cm$^{-2}$/V, where $p_{\text{top}}$ is the top surface accumulated hole density. In this region, the longitudinal sheet conductance of the Cr-TI film, $G_S = 1/R_S$ ($R_S$ is the longitudinal sheet resistance), also changes linearly, as shown in Fig. S5a. Consequently, the top surface accumulated hole mobility can be derived as $\mu_h = \frac{dG_S}{dV_g}/(e\frac{dp_{\text{top}}}{dV_g}) = 85$ cm$^2$/(V s). When the Cr-TI film is biased to the *n*-type regime, as shown in Fig. 2b region III in the main text, $c_{\text{eff}}$ almost reaches the linear region and the overall topological SS electrons (from both top and bottom surfaces) dominate the transport. Using similar argument by considering both $c_{\text{eff}}$ and $R_S$, the topological SS Dirac electron mobility can be estimated as $\mu_D = 100$ cm$^2$/(V s). When the Cr-TI film is biased to region II in Fig. 2b in the main text, $R_L$ reaches a peak at $V_g = +3$V which means the top surface carriers are mostly depleted, and meanwhile $c_{\text{eff}} = \frac{(p_{\text{bulk}}\mu_h + n_{\text{SS,b}}\mu_D)^2}{p_{\text{bulk}}\mu_h^2 - n_{\text{SS,b}}\mu_D^2}$ diverges. Here, $p_{\text{bulk}}$ is the bulk ordinary hole density (sheet density) and $n_{\text{SS,b}}$ is the bottom topological SS electron density. By combining $p_{\text{bulk}}\mu_h^2 - n_{\text{SS,b}}\mu_D^2 = 0$ and $e(p_{\text{bulk}}\mu_h + n_{\text{SS,b}}\mu_D) = 1/R_{S,\text{peak}}$, we find that $n_{\text{SS,b}} = 1.02 \times 10^{12}$ cm$^{-2}$ and $p_{\text{bulk}} = 1.4 \times 10^{12}$ cm$^{-2}$. Here, $R_{S,\text{peak}}$ is the sheet resistance at $V_g = +3$V and in our device geometry (L=W), $R_{S,\text{peak}} = R_{L,\text{peak}}$. Based on the simple capacitor model and the carrier-attracting rate ($9 \times 10^{10}$ cm$^{-2}$/V) by the top gate, the top surface carrier density under different gate voltages $V_g$ can be obtained, as plotted in the upper panel of Fig.4 in the main text. We find that the bottom topological SS electron density $n_{\text{SS,b}}$ is always larger than the top topological SS electron density $n_{\text{SS,t}}$ in the whole n-type regime, which is consistent with the two surface carrier distributions in the non-doped TI film under different gate voltages as discussed above.

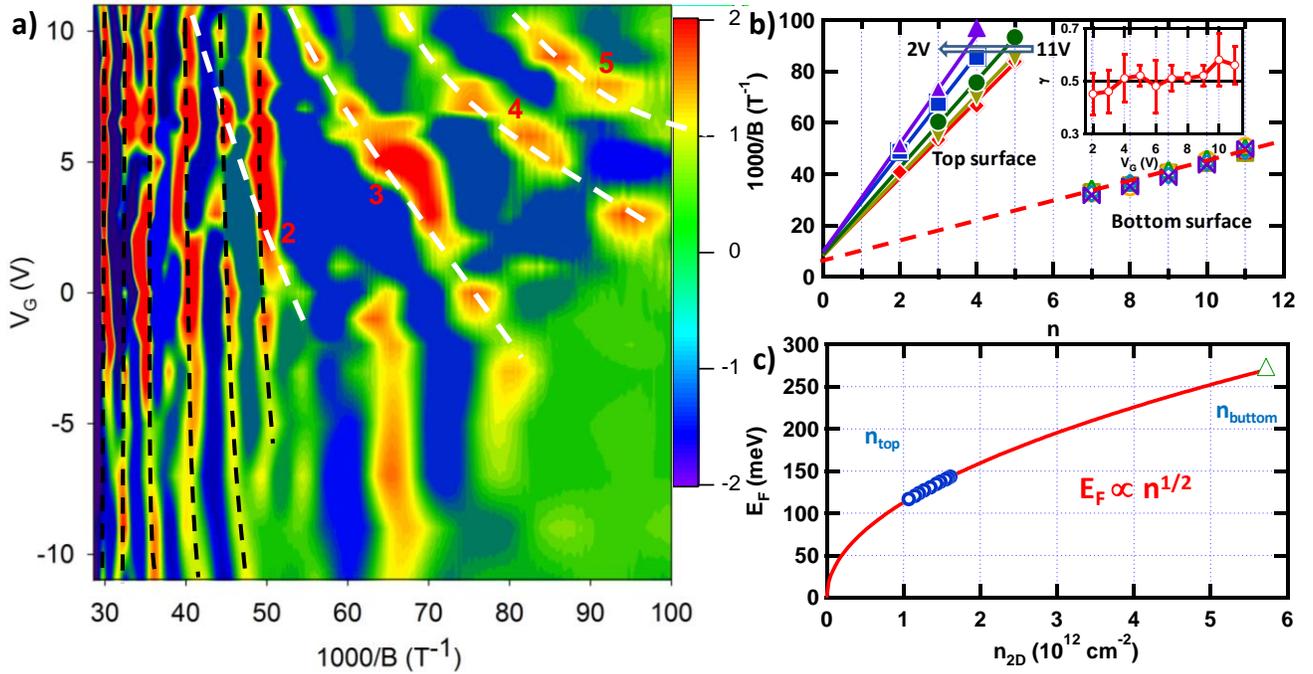

**Figure S4. SdH quantum oscillations from the top and bottom surface states in the $(Bi_{0.53}Sb_{0.47})_2Te_3$ thin film.** (**a**) $d^2R_L/dB^2$ as a function of $1/B$ and $V_g$. Both gate dependent and independent peaks are observed. The peaks which change with $V_g$ originate from the formation of Landau levels of Dirac fermions on the top surface (white dashed lines, Landau levels 2 to 5 are marked). The $V_g$-independent peaks come from the formation of Landau levels of the bottom surface Dirac fermions (black dashed lines). (**b**) Landau fan diagram of the peaks. The peaks of the top surface Landau levels (solid symbols) show systematic change depending on the gate voltage, while those of the bottom surface Landau levels are almost constant. Inset: The intercept $\gamma$ as a function of the gate voltage. The black line indicates $\gamma = 0.5$. (**c**) The carrier density of the top (circles) and bottom (triangle) surface states as a function of Fermi level $E_F$ extracted from the corresponding SdH oscillations for various gate voltages. A quadratic relationship is shown. This figure is adapted with permission from ref. 4. Copyright (2013) Nature Publishing Group.

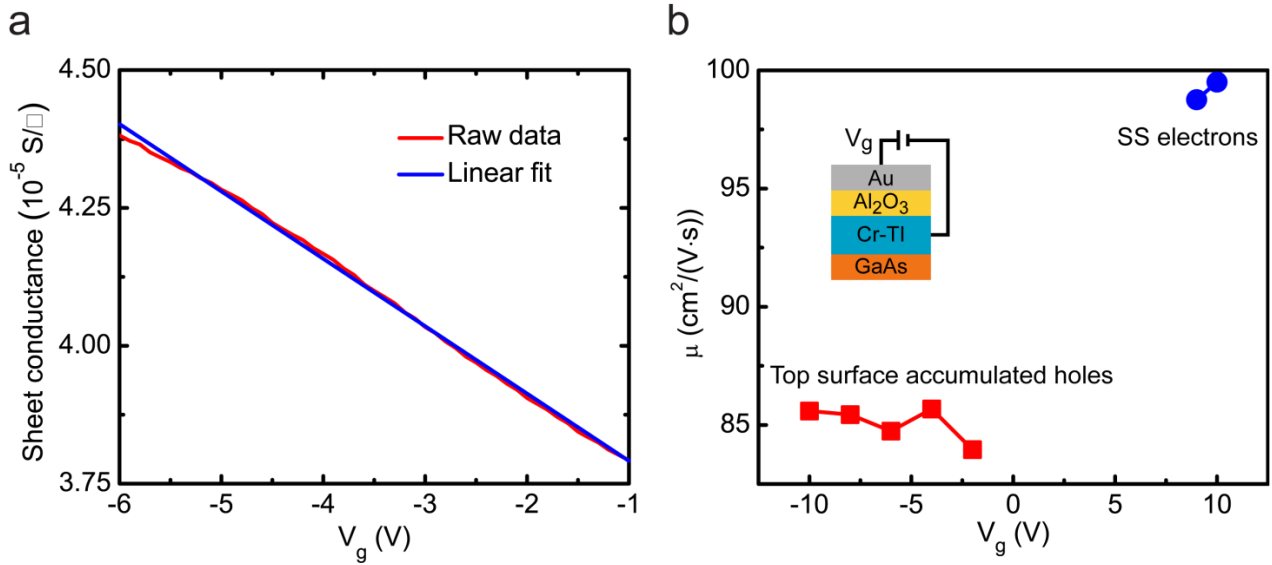

**Figure S5. a**, The longitudinal sheet conductance $G_S$ as a function of the gate voltage $V_g$ in the *p*-type regime in the Au(electrode)/$Al_2O_3$/Cr-TI/GaAs Hall bar structure. **b**, The extracted surface carrier mobilities as functions of the gate voltage $V_g$ in both *p* and *n*-type regimes in the Au(electrode)/$Al_2O_3$/Cr-TI/GaAs Hall bar structure. The transport experiments were performed at 1.9K. Inset: schematic of the structure and the applied gate voltage.

## 4. Determination of the out-of-plane anisotropy coefficient $K$ and the saturation AHE resistance $R_A$

We use the rotation experiment [11] to extract the out-of-plane anisotropy coefficient $K$ and the saturation AHE resistance $R_A$. In the rotation experiment, we apply a large external magnetic field of constant magnitude, $B_{ext} = |\boldsymbol{B}_{ext}| = 2T$, to ensure single domain behavior. Then we continuously rotate the orientation of the external magnetic field in the $yz$-plane while simultaneously measuring the Hall resistance $R_H$. This $R_H$ is composed of two parts, $R_H = \alpha B_{ext}\cos\theta_B + R_A\cos\theta_M$, where $\alpha$ is the out-of-plane ordinary Hall slope and $R_A$ is the out-of-plane saturation AHE resistance; $\theta_B$ and $\theta_M$ are the polar angles of $\boldsymbol{B}_{ext}$ and $\boldsymbol{M}$ (the magnetization) with respect to the $z$-axis, respectively, as shown in Fig. S6 inset. Here, we define $\theta_B = 0$ when $\boldsymbol{B}_{ext}$ is pointing along $z$-axis and the clockwise direction as the positive rotation direction. The ordinary Hall resistance part can be subtracted from $R_H$ by measuring the ordinary Hall slope $\alpha$ at the large field region when sweeping the out-of-plane magnetic field. The part left is the AHE resistance $R_{AHE} = R_A\cos\theta_M$. The relation between $\theta_M$ and $\theta_B$ can be established by balancing the torques due to the external magnetic field and the perpendicular anisotropy field, *i.e.*, the total torque $\boldsymbol{\tau}_{tot} = -\gamma \boldsymbol{M} \times (\boldsymbol{B}_{ext} + \boldsymbol{B}_K) = 0$, which leads to the following equation (see Fig. S6 inset):

$$K\cos\theta_M\sin\theta_M = B_{ext}(\sin\theta_B\cos\theta_M - \cos\theta_B\sin\theta_M), \quad (S4.1)$$

where $K$ is the out-of-plane anisotropy coefficient. Here, the relation $\boldsymbol{B}_K = K\cos\theta_M\hat{\boldsymbol{z}}$ is utilized since the perpendicular anisotropy field is proportional to the z-component magnetization, $M_z$. For different $K$ values, the relation between $\theta_M$ and $\theta_B$ can be established by solving equation (S4.1) numerically. Consequently, the numerically obtained AHE resistance $R_{AHE} = R_A\cos\theta_M$ can be a function of $\theta_B$. For example, in the $Al_2O_3$/Cr-TI/GaAs structure, the experimentally measured AHE resistance is plotted as a function of $\theta_B$ from the rotation experiment, as shown by the open squares in Fig. S6, where the saturation AHE resistance $R_A$ is determined to be 2.4 K$\Omega$. The solid circles in Fig. S6 represent the numerical fitting by setting $K$ to 0.6T and solving equation (S4.1). Consequently, both $R_A$ and $K$ are obtained from the fitting. The $R_A$ and $K$ in the Au(electrode)/Al$_2$O$_3$/Cr-TI/GaAs structure under different gate biases can be obtained using the same method.

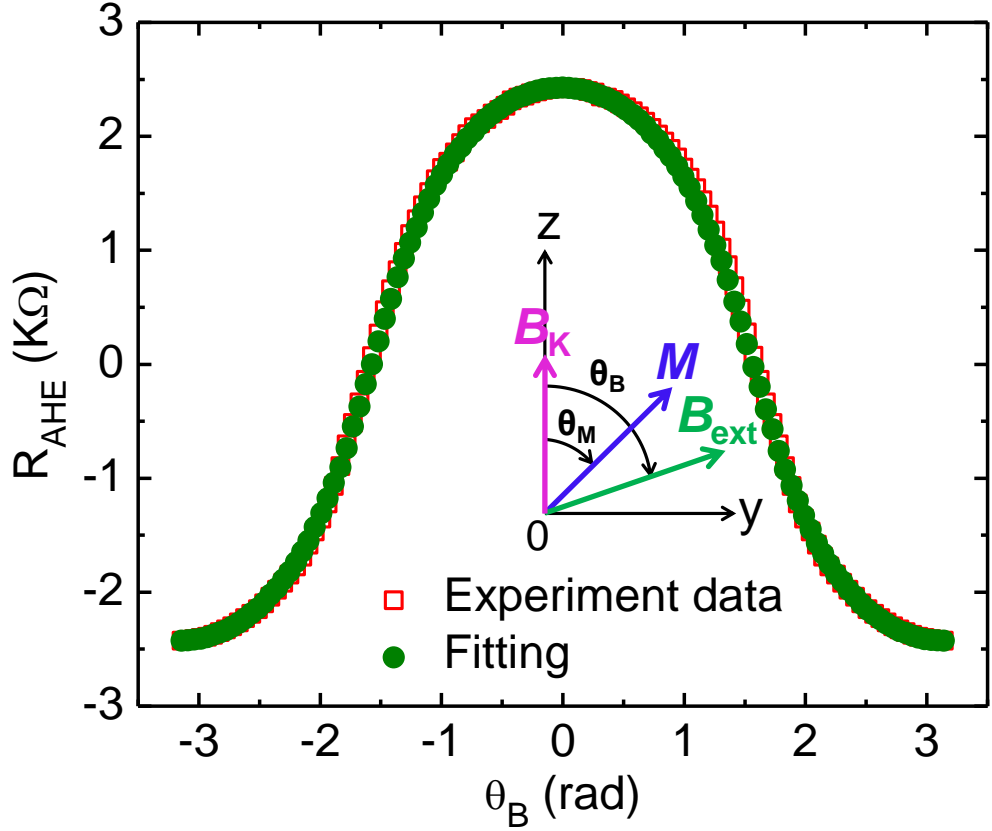

**Figure S6.** The AHE resistance $R_{AHE}$, obtained from the rotation experiment (open squares) and from the numerical fitting by solving equation (S3.1) (solid circles), as functions of the field angle $\theta_B$, respectively. In the fitting, we used $K = 0.6$T. Inset: the equilibrium orientation of the magnetization $\boldsymbol{M}$ in the presence of an external magnetic field $\boldsymbol{B}_{ext}$ with a constant magnitude of 2T. $\boldsymbol{B_K} = K\cos\theta_M \hat{z}$ is the perpendicular anisotropy field. $\theta_M$ and $\theta_B$ are the polar angles of $\boldsymbol{M}$ and $\boldsymbol{B}_{ext}$ from the z-axis, respectively. This experiment is performed at 1.9 K.

## 5. Determination of the effective spin-orbit field $B_{SO}$

In this section, within the single domain model, we derive the expression for the second harmonic AHE resistance for the case of varying magnitude of the in-plane external magnetic field applied along the $y$-axis. We are interested in the regime where the magnitude of the in-plane external magnetic field, $|B_y|$, is much larger than $K$. In this case, the equilibrium condition of the magnetization, equation (S4.1), gives $\theta_M = \pi/2$, i.e., the equilibrium orientation of the magnetization in the absence of current-induced effective spin-orbit field is pointing along the $y$-axis. However, when an AC current, $I = I_0\sin(\omega t)$, is applied, the corresponding effective spin-orbit field, $\boldsymbol{B_{SO}} = B_{SO}\sin(\omega t)\hat{\boldsymbol{z}}$, transverse to the magnetization, will cause the orientation of the magnetization to oscillate with a small amplitude $\delta\theta_M$, whose magnitude depends on the strength of the in-plane external magnetic field. This $\delta\theta_M$ gives rise to a second harmonic AHE resistance as explained in refs. 11-13. The amplitude $\delta\theta_M$, for the given $I_0$ and $\boldsymbol{B_y} = B_y\hat{\boldsymbol{y}}$, can be obtained by balancing the torques due to the external magnetic field, the anisotropy field and the current-induced effective spin-orbit field (see Fig. S7), resulting in the following equation:

$$B_{SO} - K\sin(\delta\theta_M)\cos(\delta\theta_M) = -|B_y|\sin(\delta\theta_M). \tag{S5.1}$$

We are interested when the transverse fluctuation $\delta\theta_M \ll 1$. In this case, expanding equation (S5.1) up to first order in $\delta\theta_M$ gives the solution, $\delta\theta_M = -B_{SO}/(|B_y| - K)$. In our measurement, the second harmonic AHE resistance is defined as [11], $R^{2\omega}_{AHE} = -\frac{1}{2}\frac{dR_{AHE}}{dI}I_0$, and in this special case, $R^{2\omega}_{AHE} = -\frac{R_A}{2}\frac{d\cos\theta_M}{dI}I_0 = \frac{R_A}{2}\frac{\delta\theta_M}{\delta I}I_0$. Since $\delta I$ is just $I_0$, the second harmonic AHE resistance can be readily derived:

$$R^{2\omega}_{AHE} = -\frac{1}{2}\frac{R_A B_{SO}}{(|B_y| - K)}. \tag{S5.2}$$

As shown in Figs. 1e and 3a in the main text, within the current-induced spin-orbit torque model we obtained and verified the $1/(|B_y| - K)$ dependence of the second harmonic AHE signal.

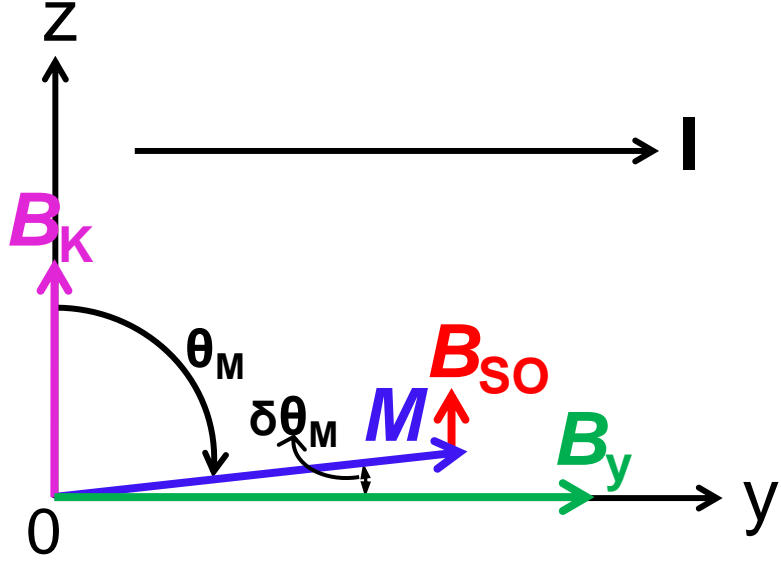

**Figure S7.** Illustration of the magnetization orientation in the presence of a large in-plane external magnetic field $B_y$ along the $y$-axis and an AC current, $I = I_0\sin(\omega t)$. $B_{SO}$ is the current-induced effective spin-orbit field; $\delta\theta_M$ is the small transverse deflection angle caused by $B_{SO}$. $B_K = K\cos\theta_M \hat{z}$ is the perpendicular anisotropy field. $\theta_M \approx \pi/2$.

## 6. Magnetization magnitude as a function of the gate voltage $V_g$

In order to estimate the magnetization magnitude $M_S$ under different gate voltages $V_g$, we first measured the Curie temperature $T_C$ of the Cr-TI layer in the Au(electrode)/Al$_2$O$_3$/Cr-TI/GaAs structure under different gate voltages $V_g$ using the same method as discussed in Section 1 (Fig. S1c). $T_C$ shows a monotonic decrease as $V_g$ increases from -10V to +10V, as displayed in Fig. S8a, which is consistent with our previous studies [1,2] and in accordance with the out-of-plane anisotropy coefficient change with respect to $V_g$ in the main text (Fig. 2d). Meanwhile, within the accessibility of our superconducting quantum interference device (SQUID) measurement system, we carried out the magnetization versus magnetic field measurement at 5K, 8K, 10K and 11K when no gate voltage was applied on the sample. The magnetization magnitude $M_S$ can be obtained after subtracting the linear background at large field, and the results are plotted in Fig. S8b as the solid squares. It can be seen that $M_S$ is almost 0 at $T = 11$K, consistent with the Curie temperature at $V_g = 0$V (Fig. S8a). Furthermore, it is known that the $M_S - T$ relation follows the power law [14] $M_S = M_0(1 - \frac{T}{T_C})^\beta$ in near the Curie temperature region. Subsequently, we carried out the fitting using the measured data, as shown by the purple curve ($V_g = 0$V) in Fig. S8b, and find $M_0 = 9.6$ emu/cm$^3$, $\beta = 0.6$. After that, we use the same power law formula but different $T_C$ values as presented in Fig. S8a to estimate the $M_S - T$ relation under various gate voltages. The obtained results are plotted in Fig. S8b. It can be seen that at $T = 1.9$K, which is the experimental temperature we mostly used in the main text, the magnetization magnitude $M_S$ does not change much. In Fig. S8c, we plot the $M_S$, obtained from the fittings in Fig. S8b, versus the gate voltage $V_g$ for $T = 1.9$K. We can observe that $M_S$ changes from 8.6 emu/cm$^3$ ($V_g = -10$V) to 8.3 emu/cm$^3$ ($V_g = +10$V), which is modulated by less than 5% and indicates that the change in magnetization magnitude $M_S$ is almost negligible in the gate-controlled experiments carried out in the main text.

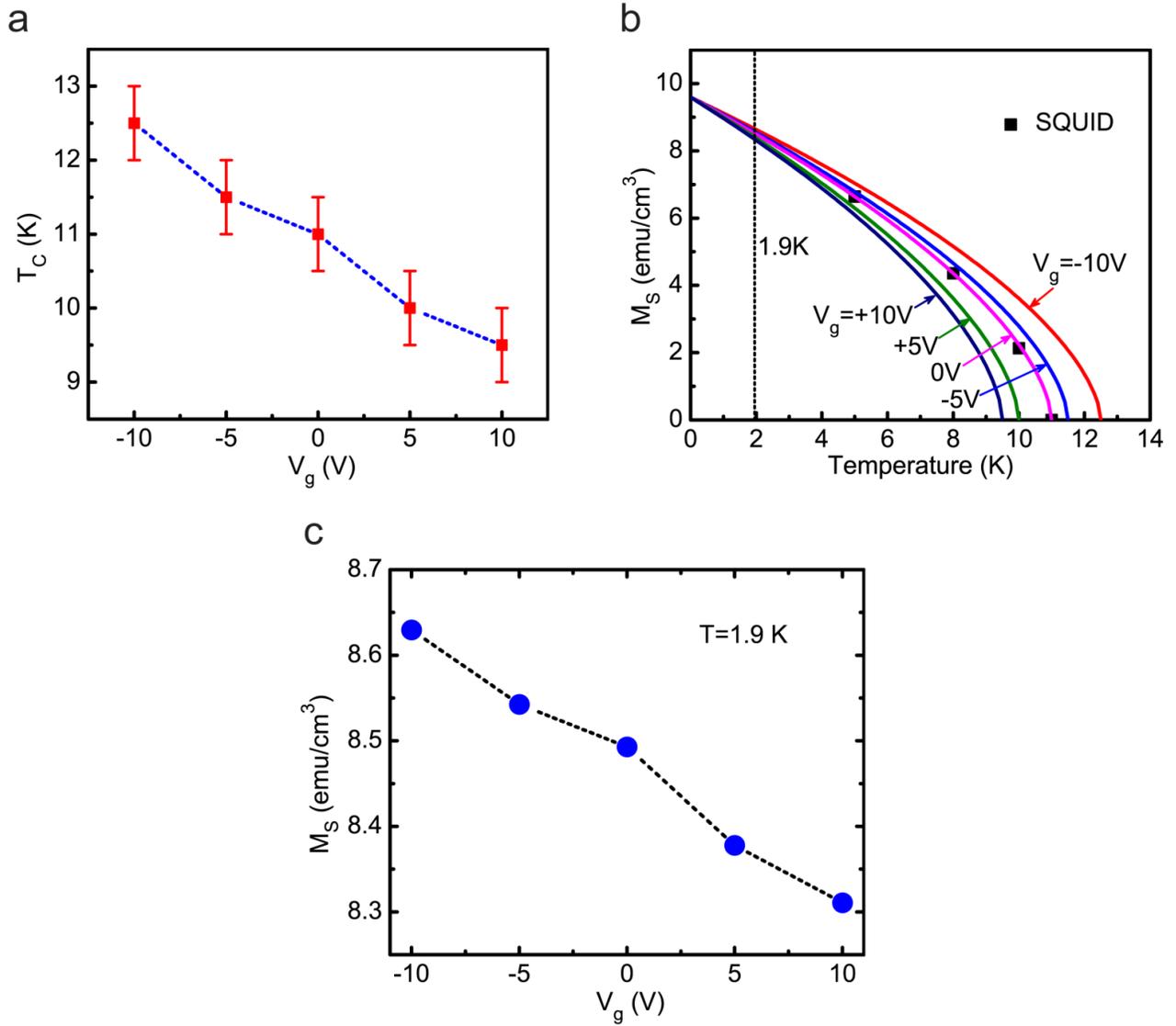

**Figure S8. a**, Curie temperature $T_C$ as a function of gate voltage $V_g$ in the Au(electrode)/Al$_2$O$_3$/Cr-TI/GaAs structure. **b**, Solid squares represent the magnetization magnitude $M_S$ measured from SQUID at different temperatures when no gate voltage is applied. Different curves show the theoretical fittings of the $M_S - T$ relation under different gate voltages. Dashed line denotes $T = 1.9$ K. **c**, Magnetization magnitude $M_S$ obtained from the fitting for $T = 1.9$ K as a function of the gate voltage $V_g$.